 \documentclass[twocolumn]{aastex631}

\shorttitle{Superbright novae in our Galaxy}
\shortauthors{Hachisu \& Kato}

\begin{document}

\title{Identification of a superbright nova in our Galaxy:
revisiting the light curve analyses of the candidate novae
CP Lac, CP Pup, V838 Her, V597 Pup, V5583 Sgr, and V5589 Sgr}


\author[0000-0002-0884-7404]{Izumi Hachisu}
\affil{Department of Earth Science and Astronomy,
College of Arts and Sciences, The University of Tokyo,
3-8-1 Komaba, Meguro-ku, Tokyo 153-8902, Japan}
\email{izumi.hachisu@outlook.jp}

\author[0000-0002-8522-8033]{Mariko Kato}
\affil{Department of Astronomy, Keio University,
Hiyoshi, Kouhoku-ku, Yokohama 223-8521, Japan}




.


\begin{abstract}
The two very fast novae in our Galaxy, V1500 Cyg and V1674 Her, reached
the absolute $V$ magnitude of $M_{V,\rm max}\approx -10.4$.
These brightest novae are identified as superbright novae.
Such superbrightnesses are realized if an optically thick shocked shell
forms far outside the nova (white dwarf) photosphere and expands
at the velocity of a few thousands km s$^{-1}$.
We have analyzed $V$ light curves of six candidates novae, 
CP Lac, CP Pup, V838 Her, V597 Pup, V5583 Sgr, and V5589 Sgr
by comparing with our model nova light curves of 
1.25 $M_\sun$ and 1.35 $M_\sun$ white dwarfs
calculated with the free-free emission from a nova wind. 
Analyzing the data with our method, we obtain $M_{V, \rm max}=-10$
for V838 Her, and we suggest that its large peak optical luminosity
is due to the ejection of an optically thick shell like
we suggested for the case of V1674 Her.
The light curves of the other five novae, 
CP Lac, CP Pup, V597 Pup, V5583 Sgr, and V5589 Sgr
are reproduced only with the free-free emission model $V$ light curves
without an optically thick shocked shell.
Their $M_{V,\rm max}$ are fainter than $-10$ mag.
We conclude that these five are not superbright novae.
\end{abstract}


\keywords{novae, cataclysmic variables ---
stars: individual (CP~Lac, CP~Pup, V5583~Sgr, V5589~Sgr, V597~Pup, V838~Her)
--- stars: winds}



\section{Introduction}
\label{introduction}

\begin{figure*}
\epsscale{1.15}
\plotone{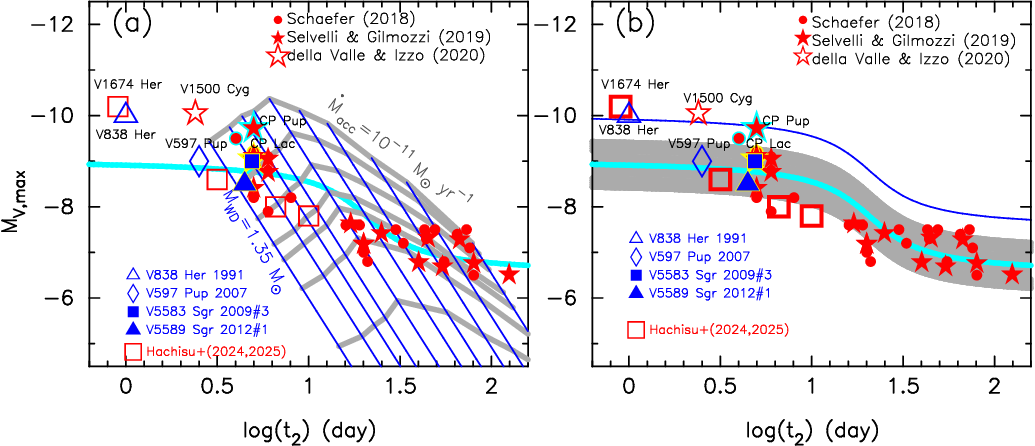}
\caption{
(a) Maximum $V$ magnitude versus rate of decline (MMRD) diagram against
the white dwarf (WD) mass and mass accretion rate to the WD.
The blue lines indicate model equi-WD mass ($M_{\rm WD}$)
lines, from left to right, 1.35, 1.3, 1.25, 1.2, 1.1, 1.0, 0.9, 0.8, 0.7,
and $0.6~M_\sun$.  The thick solid gray lines denote model equi-mass
accretion rate ($\dot M_{\rm acc}$) lines, from lower to upper, 
$3\times 10^{-8}$, $1\times 10^{-8}$, $5\times 10^{-9}$, $3\times 10^{-9}$,
$1\times 10^{-9}$, $1\times 10^{-10}$, and $1\times 10^{-11} M_\sun$~yr$^{-1}$.
These lines are taken from \citet{hac20skhs} based on the optically thick
nova wind model \citep{kat94h}
and nuclear runaway model calculation of mass accretion onto each WD.
The brightnesses of novae are calculated from free-free emission 
luminosity of Equation (\ref{free-free_flux_v-band}).
We overplot filled red circles taken from ``Golden sample'' of
\citet{schaefer18}, filled stars from \citet{sel19},
and open star (V1500~Cyg) from \citet{del20i}.
CP Lac and CP Pup are denoted by red symbols encircled
by yellow and cyan lines, respectively.
The four novae (KT Eri, V339 Del, V392 Per, and V1674 Her) are
taken from \citet{hac25kw}, \citet{hac24km}, \citet{hac25kv392per},
and \citet{hac26kv1674her3}, respectively.
The thick solid cyan line indicates the empirical
line for the MMRD relation obtained by \citet{del20i}.
The positions of four novae, V838 Her, V597 Pup, V5583 Sgr, and V5589 Sgr
are taken from the present work in advance.
(b) Same as panel (a), but without theoretical model lines. 
The gray region notes $\pm 0.5$ mag area around the thick cyan line, 
of which the 1 mag shifted up is depicted by the blue line. 
\label{max_t2_selvelli2019_schaefer2018_hachisu_plus_saio_2fig}}
\end{figure*}

A classical nova is a thermonuclear explosion of a hydrogen-rich envelope
on a mass-accreting white dwarf (WD) in a binary.  Hydrogen ignites to
trigger an outburst when the mass of the envelope reaches a critical value
\citep[e.g.,][]{nar80, ibe82, pri95k, sio79, spa78}.
The WD photosphere expands to a giant star size and blows massive winds
 \citep[e.g.,][for recent fully self-consistent nova explosion
models]{kat22sha, kat25hsa}.  Radioactive $^{7}$Be lines,
observed  by \citet{taj15sn} in the 2013 outburst of V339 Del,
are the first direct evidence of thermonuclear runaway event
of a nova \citep[see, e.g.,][for a commentary]{her15}.
The optical brightness of a nova typically reaches the absolute $V$
magnitude of $M_V\sim -8\pm 1$ and sometimes up to $M_V\sim -10$
(Figure \ref{max_t2_selvelli2019_schaefer2018_hachisu_plus_saio_2fig}).

\subsection{Brightness distribution of novae}
\label{brightness_distribution}
The peak absolute $V$ brightness $M_{V,\rm max}$ of a nova is sometimes related
to its decline rate, $t_2$ or $t_3$, where $t_2$ and $t_3$ are
the days during which the $V$ magnitude decays by 2 and 3 mag
from the $V$ peak, respectively.
The $M_{V,\rm max}$ versus $t_2$ diagrams for many novae
in our Galaxy are shown in Figure
\ref{max_t2_selvelli2019_schaefer2018_hachisu_plus_saio_2fig}.
The peak brightness of a nova outburst depends mainly on the WD mass
($M_{\rm WD}$) and mass accretion rate ($\dot{M}_{\rm acc}$) to the WD,
so that  Figure
\ref{max_t2_selvelli2019_schaefer2018_hachisu_plus_saio_2fig}(a) shows
a model peak brightness against the WD mass and mass accretion rate, 
the data of which are taken from \citet{hac20skhs}.  See \citet{hac20skhs}
for the other weaker parameter dependences.
In general, a smaller $t_2$ nova shows a brighter $V$ peak ($M_{V,\rm max}$).
The statistical mean trend between $M_{V,\rm max}$ and $\log t_2$ (or
$\log t_3$) is called the Maximum Magnitude versus Rate of Decline (MMRD)
relation \citep[e.g., ][]{mcl45, ros64, cap89dd, dow00, shaft11dh,
schaefer18, sel19, del20i}.

We plot one of such MMRD relations by the thick cyan line in
Figure \ref{max_t2_selvelli2019_schaefer2018_hachisu_plus_saio_2fig}, i.e.,
\begin{equation}
M_{V,\rm max} = -7.78-0.81 \times \arctan\left( 
{{1.32 - \log t_2}\over {0.23}}\right).
\label{rev_s_shape_mmrd_relation}
\end{equation}
This type of MMRD relations are called the reverse S-shape MMRD relations
\citep[e.g.,][]{del20i}.
The absolute $V$ brightness $M_V$ of a nova is calculated to be
\begin{equation}
M_V = m_V - (m-M)_{V, \rm nova},
\label{absolute_magnitude_v}
\end{equation}
where $m_V$ is the apparent $V$ magnitude and $(m-M)_{V, \rm nova}$
is the distance modulus in the $V$ band toward a specific nova,
which is related to the distance $d$ and extinction $E(B-V)$ by
\begin{eqnarray}
(m-M)_{V, \rm nova} &=& 5 \log \left( {d \over {\rm 10~pc}} \right) + A_V \cr
 &=& 5 \log \left( {d \over {\rm 10~pc}} \right) + 3.1 E(B-V),
\label{distance_modulus_extinction}
\end{eqnarray}
where $A_V$ is the absorption in the $V$ band.

Many novae follow the thick cyan line of Equation 
(\ref{rev_s_shape_mmrd_relation}) within a scatter of $\Delta V=\pm 0.5$ mag
(gray shadow in Figure 
\ref{max_t2_selvelli2019_schaefer2018_hachisu_plus_saio_2fig}(b)).
Such a relation has long been used to estimate the distance
modulus in the $V$ band, $\mu_V\equiv (m-M)_V$, and distance to a nova
if $t_2$ is obtained from the $V$ light curve of a nova.
However, there are novae that largely deviate from this MMRD line,
so the MMRD relation should not be used to estimate the distance to a nova.
From the theoretical point of view, $M_{V, \rm max}$ depends mainly
on the WD mass and mass-accretion rate to the WD.
Therefore, the $M_{V, \rm max}$ cannot be uniquely determined
by the decline rate of $t_2$ (or $t_3$), as shown in Figure
\ref{max_t2_selvelli2019_schaefer2018_hachisu_plus_saio_2fig}(a).

\subsection{Superbright novae}
\label{super_bright_novae}
In Figure \ref{max_t2_selvelli2019_schaefer2018_hachisu_plus_saio_2fig}(b),
many novae are located around ($\pm 0.5$ mag, gray-shaded) the thick cyan
line of Equation (\ref{rev_s_shape_mmrd_relation}), while a few novae are
$\gtrsim 1$ mag (blue line) brighter than the thick cyan line.
\citet{del91} dubbed them ``super-bright novae'' that reached the absolute
$V$ brightness of $M_{V, \rm max} \lesssim -10$ mag and $\gtrsim$1 mag
brighter than a typical MMRD line \citep[similar to the thick cyan line in
Figure \ref{max_t2_selvelli2019_schaefer2018_hachisu_plus_saio_2fig},
but he had taken from ][]{cap89dd}.
\citet{del91} cited the classical nova V1500 Cyg as a superbright nova
in our Galaxy.  These superbright novae are located outside 
\citet{hac20skhs}'s theoretical region of nova MMRD distribution
as explained below in Section \ref{free-free_emission_light_curves} and 
in Figure \ref{max_t2_selvelli2019_schaefer2018_hachisu_plus_saio_2fig}(a).




\subsection{Free-free emission nova light curves}
\label{free-free_emission_light_curves}

From the theoretical point of view,  the peak magnitude $M_{V, \rm max}$ 
is not uniquely determined by the decline rate of $t_2$ (or $t_3$).  
It is not expressed by a single line like the thick cyan line in 
Figure \ref{max_t2_selvelli2019_schaefer2018_hachisu_plus_saio_2fig}. 
Because a nova explosion depends mainly on the WD mass ($M_{\rm WD}$)
and mass accretion rate ($\dot{M}_{\rm acc}$) to the WD, the peak magnitude
$M_{V, \rm max}$ and the decline rate of $t_2$ (or $t_3$) depends
differently on these values.  Thus, the relation between them cannot
be expressed by a single line.

\citet{hac20skhs} calculated a number of post-maximum nova
light curves with different sets of the WD mass and mass accretion rate.  
They calculated the decay phase of each nova based on the optically
thick wind theory \citep{kat94h},
assuming that the optical flux is dominated by free-free emission
\citep[e.g.,][]{enn77,gal76}, which can be simplified as
\citep{hac06kb,hac20skhs} 
\begin{equation}
L_{V, \rm ff,wind} = A_{\rm ff} ~{{\dot M^2_{\rm wind}}
\over{v^2_{\rm ph} R_{\rm ph}}},
\label{free-free_flux_v-band}
\end{equation}
where $L_{\rm V, \rm ff,wind}$ is the free-free emission luminosity in
the $V$ band, $\dot{M}_{\rm wind}$ is the wind mass loss rate, $v_{\rm ph}$ 
is the velocity at the photosphere, $R_{\rm ph}$ is the photospheric radius
of a nova wind, and $A_{\rm ff}$ is the coefficient for a nova 
\citep[see ][for their calculations]{hac20skhs, kat25hsa}.

We depict such an MMRD ($M_{V,\rm max}$ versus $\log t_2$) 
diagram against the equi-WD mass lines and equi-mass accretion rate lines
in Figure \ref{max_t2_selvelli2019_schaefer2018_hachisu_plus_saio_2fig}(a),
taken from \citet{hac20skhs}.
Almost all of novae fall in \citet{hac20skhs}'s theoretical region
except the three novae, V1500 Cyg 1975, V838 Her 1991, and V1674 Her 2021,
the reason of which is the central topic of this paper.

\begin{deluxetable*}{lllrcllllllc}
\tabletypesize{\scriptsize}
\tablecaption{List of candidate superbright novae in our Galaxy
\label{table_brightness_novae}}
\tablehead{
\colhead{Nova} &
\colhead{$M_{\rm WD}$} &
\colhead{$\dot M_{\rm acc}$} &
\colhead{$t_{\rm rec}$} &
\colhead{$(m-M)_V$} &
\colhead{$t_2$\tablenotemark{a}} &
\colhead{$M_{V,\rm max}$} &
\colhead{$E(B-V)$} &
\colhead{$d$}&
\colhead{$d_{\rm Gaia}$\tablenotemark{b}}&
\colhead{$M_{\rm WD}$\tablenotemark{c}} &
\colhead{superbright}\\
\colhead{ }&
\colhead{($M_\sun$)} &
\colhead{($M_\sun$ yr$^{-1}$)} &
\colhead{(yr)}&
\colhead{ }&
\colhead{(day)}&
\colhead{ } &
\colhead{ } &
\colhead{(kpc)}& 
\colhead{(kpc)}& 
\colhead{($M_\sun$)} &
\colhead{ }
}
\startdata
V1500 Cyg &1.25 & $5\times 10^{-11}$ & 120,000 & 12.3 & 2  & $-10.4$ & 0.43 & 1.56  & $1.57_{-0.19}^{+0.27}$ & 1.2 & yes \\ 
V1674 Her &1.35 & $1\times 10^{-11}$ & 160,000 & 16.3 & 0.9 & $-10.2$ & 0.5 & 8.9 & $6.0_{-2.8}^{+3.8}$ & 1.35 & yes \\
V838 Her &1.35 & $1\times 10^{-11}$ & 160,000 & 15.3  & 1  & $-10.0$ & 0.38  & 6.7 & $5.5_{-1.8}^{+1.9}$ & 1.35 & yes \\
CP Lac &1.25 & $1\times 10^{-9}$ & 4,000 & 11.0  & 5  & $-9.1$ & 0.24 & 1.1 & $1.13_{-0.03}^{+0.03}$ & 1.25 & no \\ 
CP Pup &1.25 & $5\times 10^{-11}$ & 120,000 & 10.0  & 5  & $-9.6$ & 0.2 & 0.76 & $0.76_{-0.01}^{+0.01}$ & 1.25 & no \\ 
V597 Pup &1.35 & $1\times 10^{-11}$ & 160,000 & 15.8  & 2.5  & $-9.0$ & 0.38 & 8.4 & $4.8_{-2.8}^{+2.0}$ & 1.35 & no \\ 
V5583 Sgr &1.25 & $5\times 10^{-10}$ & 8,100 & 16.4  & 5  & $-9.0$  & 0.34 & 11.7 & \nodata & 1.23 & no \\ 
V5589 Sgr &1.35 & $5\times 10^{-10}$ & 1,900 & 17.3  & 4.5  & $-8.5$ & 0.865 & 8.4 & $6.2_{-2.1}^{+2.2}$ & 1.33 & no
\enddata
\tablenotetext{a}{$t_2$ are taken from Table 5 or 6 of \citet{ozd18} if
otherwise specified.}
\tablenotetext{b}{The Gaia DR3 geometric distance estimates are
taken from \citet{bai21rf}.}
\tablenotetext{c}{these WD masses are estimated by the steady-state envelope
evolution models of nova winds in the decline phase: V1500 Cyg \citep{hac14k},
V1674 Her \citep{kat25hsa}, V838 Her \citep{kat09v838her}, CP Lac (present
work), CP Pup (present work), V597 Pup (present work), V5583 Sgr
\citep{hac19kb}, V5589 Sgr \citep{hac19kb}.}
\end{deluxetable*}

\subsection{Formation of an optically thick shocked shell}
\label{optically_thick_shock}

Hard X-ray and GeV gamma-ray emissions have been sometimes
observed in classical novae.
Hard X-rays were detected in an intermediate phase of a nova outburst
\citep[e.g.,][]{llo92ob, bal98ko, muk01i}.
GeV gamma-ray emissions were observed in an early phase of a nova outburst,
just from the post-maximum phase, and lasted a few tens of days
\citep[e.g.,][]{abd10, ack14aa, li17mc, gor21ap}.

These high-energy (hard X-ray and GeV gamma ray)
emissions are considered to originate from strong shocks
between shells ejected with different velocities
\citep{cho14ly, met15fv, mar18dj}.
If the inner shell (later ejected) has a larger velocity than that of
the outer shell (earlier ejected), the inner one can catch up with the
outer one and forms a shock wave \citep[e.g.,][]{muk19s, ayd20ci, ayd20sc}.
The assumption of multiple shell ejection is the key of this idea.

Such a multiple shell ejection was suggested
from both optical and high-energy emissions from novae.
There is, however, no theoretical explanation had been presented 
until recently, that
naturally explains all these different wavelength observations
based on nova explosion models \citep[see][for a recent review]{cho21ms}.

Many numerical calculations
have been presented from the early thermonuclear runaway to the
extended phase of nova outbursts \citep[e.g.,][]{
pri92k, pri95k, epels07, sta09ih, den13hb, chen19wy, kat22sha, kat22shapjl}.
These works clarified that mass ejection is continuous,
no shock arises at the thermonuclear runaway,
and no multiple distinct mass ejection occurs.

\citet{hac22k} theoretically showed that a strong shock naturally arises
outside the WD photosphere just after the optical maximum,
based on \citet{kat22sha}'s fully self-consistent nova explosion model.
This is because the velocity of the nova wind at the WD photosphere
increases with time after the maximum expansion.
The later ejected matter has a larger
expansion velocity so that it catches up with the former ejected matter
and makes a strong shock. Thus, a shock is formed after the maximum expansion
of the WD photosphere ($=$optical maximum for the free-free emission model
light curves) and propagates far outside the WD photosphere.
This shock formation mechanism reasonably explains gamma-ray 
and hard X-ray emissions in classical novae \citep[e.g.,
YZ Ret, V339 Del, and V392 Per in][respectively]{hac23k, hac24km,
hac25kv392per}.

\citet{hac26kv1674her3} found that the shocked shell is optically thick
when a nova ejecta is as massive as $\sim 3\times 10^{-6} ~M_\sun$
in V1674 Her and V1500 Cyg \citep[see, e.g.,][for maximum ejecta
masses]{hac26kmaxej}.  They further showed that the superbrightness
($M_{V,\rm max}\lesssim -10$ mag) of a nova is realized
when the photosphere of an optically thick shocked shell emits photons
at the temperature of $T_{\rm ph}\sim$ 10000-5000 K and expands
at the velocity of $v_{\rm ph}\sim$ a few thousands km~s$^{-1}$.
Thus, an optically thick shocked shell plays an essential role 
in the light curves of the superbright novae V1500 Cyg and V1674 Her.

\subsection{Objectives}
\label{objectives_intro}

The aim of this paper is to search for other superbright novae
in our Galaxy and to confirm its origin of superbrightness.
We have already identified two superbright novae, V1500 Cyg and V1674 Her,
and have elucidated their origin of superbrightness to be photospheric
emission from an optically thick shocked shell with an expansion velocity
of a few thousands km s$^{-1}$ \citep{hac26kv1674her3}.

Here, we define a superbright nova as a nova whose (1) peak $V$ brightness
is one magnitude or more brighter than the typical fast novae given by
Equation (\ref{rev_s_shape_mmrd_relation}), and whose (2) position
in the MMRD diagram is outside \citet{hac20skhs}'s MMRD region in Figure
\ref{max_t2_selvelli2019_schaefer2018_hachisu_plus_saio_2fig}(a).
Based on the two selection rules
(a) $t_2\lesssim 5$ days and (b) enough $V$/visual data
around the optical peak, we pick up six candidate novae, CP Lac, CP Pup,
V838 Her, V597 Pup, V5583 Sgr, and V5589 Sgr.

We examine the superbrightnesses of these novae as follows:  We introduce fully
self-consistent nova outburst models calculated by \citet{kat25hsa, kat26hsa}
in Section \ref{fully_consistent_models}.  Then, in the following sections
(Sections \ref{full_v1500_cyg_1975}-\ref{full_v5589_sgr_2012_no1}),
we compare the $V$ light curve of each nova with the fully self-consistent
nova outburst models in the absolute $M_V$ magnitude. 
We obtain the distance modulus in the $V$ band, $(m-M)_V$, by direct fit
of our model light curves ($M_V$) with the observation ($m_V$). 
Then, we examine whether its origin of superbrightness
is an optically thick shocked shell or not.
Conclusions follow in Section \ref{sec_conclusion}.
We also estimate the distance modulus in the $V$ band of a nova
with the time-stretching method in Appendix \ref{time_stretching_method},
if there are no reliable distance estimates by the Gaia DR3 parallaxes
\citep{bai21rf}.

\section{Fully self-consistent nova explosion models}
\label{fully_consistent_models}

\citet{kat25hsa, kat26hsa} calculated nova outburst cycles
for a 1.25 $M_\sun$ WD with five mass accretion rates of
$5\times 10^{-11}$, $1\times 10^{-10}$, $5\times 10^{-10}$,
$1\times 10^{-9}$, and $5\times 10^{-9} ~M_\sun$ yr$^{-1}$, 
and also for a 1.35 $M_\sun$ WD with three mass accretion rates of
$1\times 10^{-11}$, $5\times 10^{-10}$,
and $5\times 10^{-9} ~M_\sun$ yr$^{-1}$, as tabulated in their Table 1.
They used their own Henyey type evolution code consistently combined with
steady state wind mass loss solutions as a surface boundary condition,
and calculated the WD structures from the center of the WD up to the
WD photosphere.
The method of their numerical calculation is explained 
in \citet{kat22sha, kat24M1213, kat25hsa, kat26hsa}.

\citet{kat25hsa, kat26hsa} presented $V$ light curves 
calculated with Equation (\ref{free-free_flux_v-band})
based on free-free emission of nova winds, and compared them with the
V1674 Her, KT Eri, V339 Del, V597 Pup, and SMC Nova 2016-10a light curves.
In what follows, we apply these fully self-consistent nova outburst 
models to the light curves of V1500 Cyg 1975, V1674 Her 2021, CP Lac 1936,
CP Pup 1942, V838 Her 1991, V597 Pup 2007, V5583 Sgr 2009\#3, and V5589 Sgr
2012\#1, and estimate their WD masses, mass-accretion rates, distance moduli
in the $V$ band, $(m-M)_V$, and so on.

The most important property in our analysis on superbright novae 
is the distance modulus in the $V$ band, $(m-M)_V$, to each nova.
We obtain $(m-M)_V$ by direct fit
of our model $V$ light curve (absolute $M_V$ magnitude)
with the observation (apparent $m_V$ magnitude). 
Thus, we obtain the absolute $M_V$ brightness of each nova.
It should be noted that we do not need (or use) the distance itself
to estimate the absolute $M_V$ magnitude of a nova.

\section{V1500 Cyg 1975}
\label{full_v1500_cyg_1975}



\begin{figure*}
\gridline{\fig{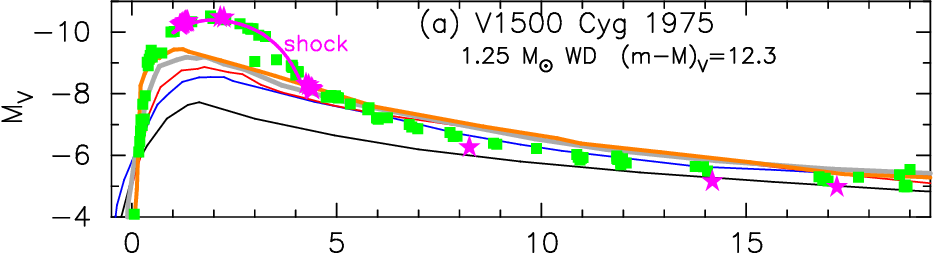}{0.75\textwidth}{}
          }
\gridline{
          \fig{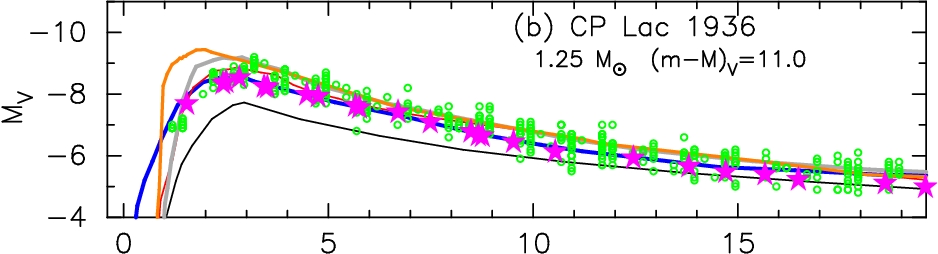}{0.75\textwidth}{}
          }
\gridline{
          \fig{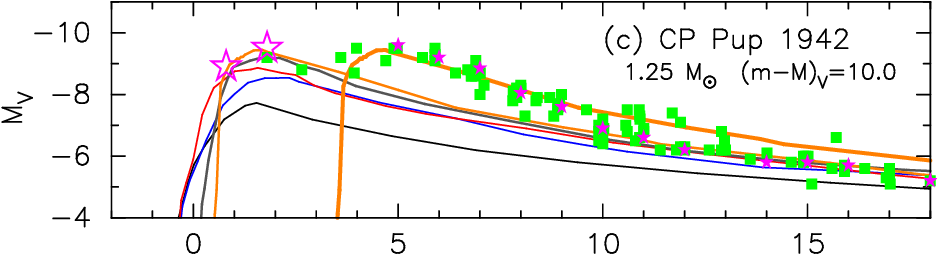}{0.75\textwidth}{}
          }
\gridline{
          \fig{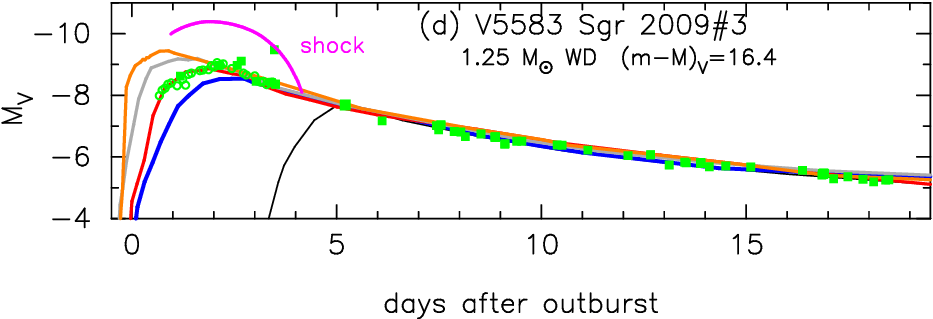}{0.75\textwidth}{}
          }
\caption{
(a) The $V$/$y$ light curves of V1500 Cyg on a linear timescale.
The data are the same as those in Figure
\ref{v1500cyg_1674her_model_v_observation_logscale}(a).
We add five free-free emission model $V$ light curves of
a 1.25 $M_\sun$ WD with
mass-accretion rates of 
$5\times 10^{-11}$ (thick orange line),
$1\times 10^{-10}$ (gray line),
$5\times 10^{-10}$ (red line),
$1\times 10^{-9}$ (blue line), and 
$5\times 10^{-9}$ (black line) $M_\sun$ yr$^{-1}$.
The thick magenta line indicates the brightness of the
optically thick shocked shell in Figure
\ref{v1500cyg_1674her_model_v_observation_logscale}(a). 
(b) CP Lac:
the green circles and magenta stars denote the visual and photoelectric
magnitudes, respectively, which are taken from AAVSO and \citet{art63k}. 
(c) CP Pup:  the data are from AAVSO (green squares),
IAUC No.925 (large unfilled magenta stars), \citet{kuip43} (magenta stars).
The thick orange line denote the same model as the thin orange line,
but is shifted by about 3 days later.
(d) V5583 Sgr:  the $V$ (filled green squares) and SMEI (small green circles) 
light curves.
\label{v1500_cyg_cp_lac_cp_pup_v5583_sgr_full_model_fit_linear}}
\end{figure*}

V1500 Cyg is an asynchronous polar of the orbital period 
$P_{\rm orb}=0.1396$ day \citep[$=3.35$ hr, e.g., ][]{pat79}.
The 1975 outburst was discovered by K. Osada
on UT 1975 August 29.48 (=JD 2442653.98) at $m_v=3.0$ (IAUC No. 2826).
Figure \ref{v1500_cyg_cp_lac_cp_pup_v5583_sgr_full_model_fit_linear}(a)
shows the $V$ (filled green squares) and $y$ (filled magenta stars)
light curves of V1500 Cyg against a linear timescale with the $V$ band
distance modulus of $(m-M)_V=12.3$ as well as
the shocked shell model light curve (magenta line labeled ``shock'')
calculated by \citet{hac26kv1674her3} and five free-free emission 
model light curves of a 1.25 $M_\sun$ WD calculated by \citet{kat26hsa}.

\citet{gal76} obtained the V1500 Cyg brightnesses for the three broad optical
$V$, $R$, and $I$ bands and the eight infrared 1.2, 1.6, 2.2, 3.6, 4.8,
8.5, 10.6, and $12.5~\mu$m bands during the 50 days following the discovery.
They concluded that the spectral energy distribution is approximately
that of a blackbody ($=$magenta line in Figure
\ref{v1500_cyg_cp_lac_cp_pup_v5583_sgr_full_model_fit_linear}(a))
during the first 3 days while it is close to
$F_\nu =$~constant after the fourth day, where $F_\nu$ is the flux
at the frequency $\nu$.  This $F_\nu =$~constant
spectra resemble those usually ascribed to the free-free emission
($=$orange line in Figure 
\ref{v1500_cyg_cp_lac_cp_pup_v5583_sgr_full_model_fit_linear}(a)).

\citet{enn77} obtained similar results, but based on the infrared
photometry from 1 to $20~\mu$m.  The nova spectrum changed from
a blackbody to a bremsstrahlung emission at day $\sim 4-5$,
that is, from that of a Rayleigh-Jeans tail ($F_\nu \propto \nu^2$)
to that of a thermal bremsstrahlung emission ($F_\nu \sim$ constant).

\citet{hac26kv1674her3} concluded that an optically thick shocked shell
expands at the velocity of $\sim 1700$ km s$^{-1}$ 
\citep{bol76g, feh76a} and emits
photons at the photosphere of recombination front in
the shocked shell from day 1.4 to day 5.
All free-free emissions from the nova winds near the WD photosphere
are absorbed by the optically thick shocked shell \citep[see Figure 3 of
 ][for a configuration of the shocked shell]{hac26kv1674her3}.
When the optically thick shocked shell becomes optically thin on day 5,
free-free emission from the nova winds outside the WD photosphere
dominates again the $V$ luminosity of V1500 Cyg.

To summarize, the nova spectrum is close to that of the blackbody around
the optical peak, and then, about 5 days after the outburst, it
enters a phase in which free-free emission dominates.
In other words, these observations can be interpreted as
the detection of the transition from optically thick photosphere
of the shocked shell (magenta line)
to free-free emission (orange line)
coming from much inner region close to the WD photosphere.
This transition accompanies a sharp drop from the $V$ peak in V1500 Cyg.


Figure \ref{v1500_cyg_cp_lac_cp_pup_v5583_sgr_full_model_fit_linear}(a)
depicts the $V$ (green squares) and $y$ (magenta stars) light curves of
V1500 Cyg in a linear timescale
together with our 1.25 $M_\sun$ WD models with five
mass-accretion rates.
Note that all the model light curves are depicted in the absolute $V$  
magnitude. The observational $V$/$y$ data are all converted to the 
absolute $V$/$y$ magnitudes with the distance modulus of $(m-M)_V=12.3$.

\begin{figure*}
\epsscale{0.75}
\plotone{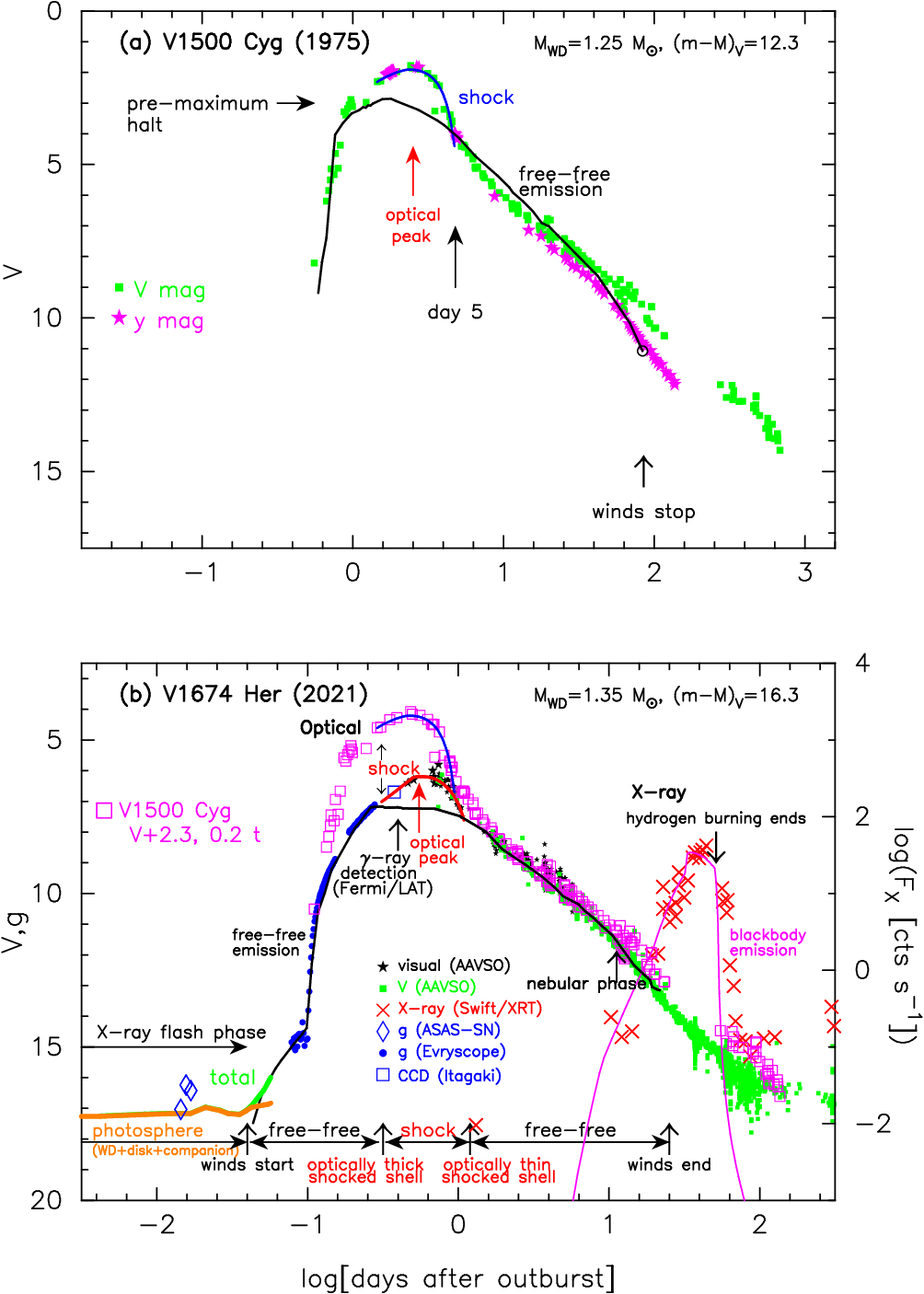}
\caption{
(a) The $V$ and $y$ light curves of V1500 Cyg are plotted against
a logarithmic time, days after outburst.  The outburst day is assumed to
be $t_{\rm OB}=$JD 2442653.0 $=$ UT 1975 August 28.5.  The $V$ data
(filled green squares) are taken from \citet{tem79} while the $y$ magnitudes
(filled magenta stars) are from \citet{lock76m}.
The free-free emission model light curve (black line)
of a $1.25 ~M_\sun$ WD with $\dot{M}_{\rm acc}=5\times 10^{-11} 
~M_\sun$ yr$^{-1}$ is taken from \citet{kat26hsa} for the distance modulus
in the $V$ band of $(m-M)_V=12.3$.
The shocked shell model light curve (blue line labeled shock) is taken from
\citet{hac26kv1674her3}, starting from the initial radius of the shock
$R_{\rm ph,sh}= 280 ~R_\sun$ on day 1.4 with
the expansion velocity of $v_{\rm shell}= 1700$ km s$^{-1}$.
See the main text in Section \ref{full_v1500_cyg_1975} for more details.
(b) Same as those in panel (a), but for the $V$, visual, $g$, and X-ray
light curves of V1674 Her.  We overlap the $V$ light curves of V1500 Cyg
(open magenta squares)
with that of V1674 Her by shifting them 2.3 mag down and 5 times squeeze
of time as denoted by ``V1500 Cyg V+2.3, 0.2 t.''  The shocked shell
light curve model (red line labeled shock) is taken from 
\citet{hac26kv1674her3}, starting from the initial radius of the shock
$R_{\rm ph,sh}= 210 ~R_\sun$ on day 0.3 with
the expansion velocity of $v_{\rm shell}= 4000$ km s$^{-1}$.
The black line indicates the free-free $V$ luminosity calculated with
Equation (\ref{free-free_flux_v-band}) based on \citet{kat25hsa}'s 1.35
$M_\sun$ WD model with $\dot{M}_{\rm acc}=1\times 10^{-11} ~M_\sun$ yr$^{-1}$.
The magenta line is its soft X-ray (0.3-10 keV) luminosity calculated with
the blackbody approximation.
\label{v1500cyg_1674her_model_v_observation_logscale}}
\end{figure*}

Among the five model $V$ light curves, we adopt the thick orange line
of $\dot{M}_{\rm acc}=5\times 10^{-11} ~M_\sun$ yr$^{-1}$.
Although the rising and decay phase are well reproduced
with the free-free emission model light curve (thick orange line),
the observed optical peak is much brighter than the model light curve.
\citet{hac26kv1674her3} calculated the $V$ light curve of an optically
thick shocked shell, which is shown by the magenta line 
labeled ``shock'' in Figure
\ref{v1500_cyg_cp_lac_cp_pup_v5583_sgr_full_model_fit_linear}(a), 
and also shown by the blue line in Figure
\ref{v1500cyg_1674her_model_v_observation_logscale}(a).
The combination of the optically thick shocked shell and
free-free emission models reasonably reproduces
the early light curve of V1500 Cyg.

We confirm the best-fit distance modulus of $(m-M)_V=12.3\pm 0.2$
when we increase $(m-M)_V$ by a 0.1 mag step
from $(m-M)_V=$11.3 to 13.3 in
Figure \ref{v1500_cyg_cp_lac_cp_pup_v5583_sgr_full_model_fit_linear}(a)
and, in each step, directly fit our model $V$ light curve
(thick orange line and magenta line) with the observation.



\begin{figure*}
\gridline{\fig{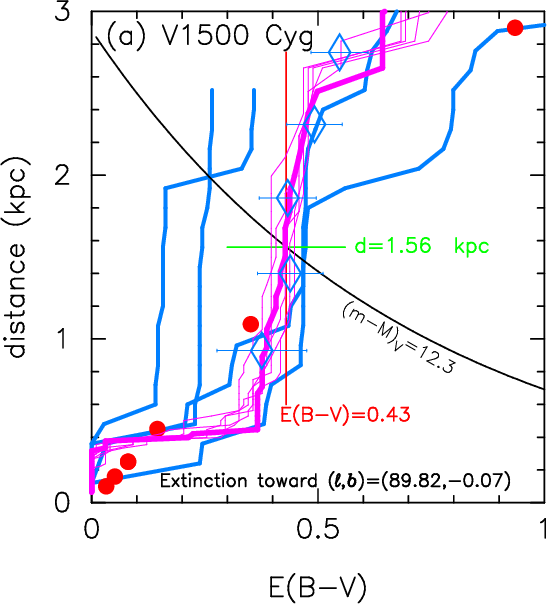}{0.4\textwidth}{}
          \fig{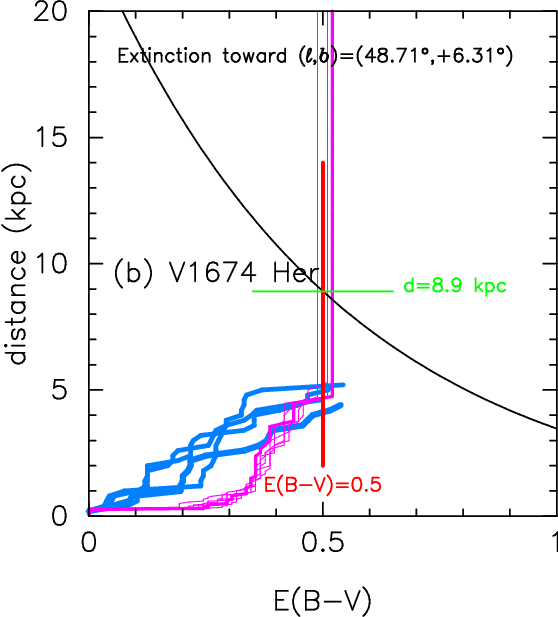}{0.4\textwidth}{}
          }
\gridline{
          \fig{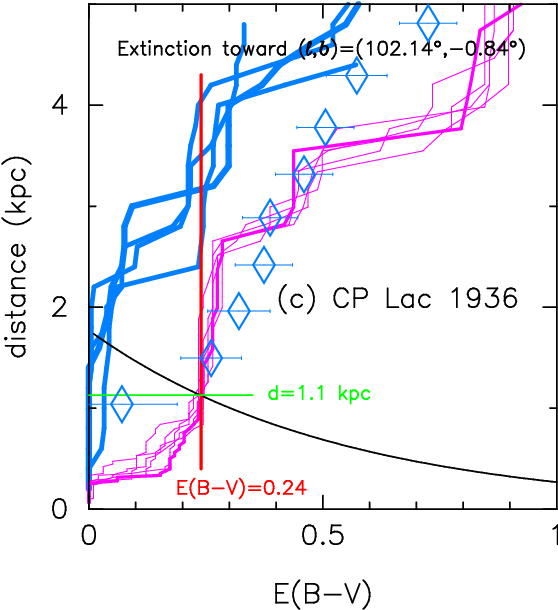}{0.4\textwidth}{}
          \fig{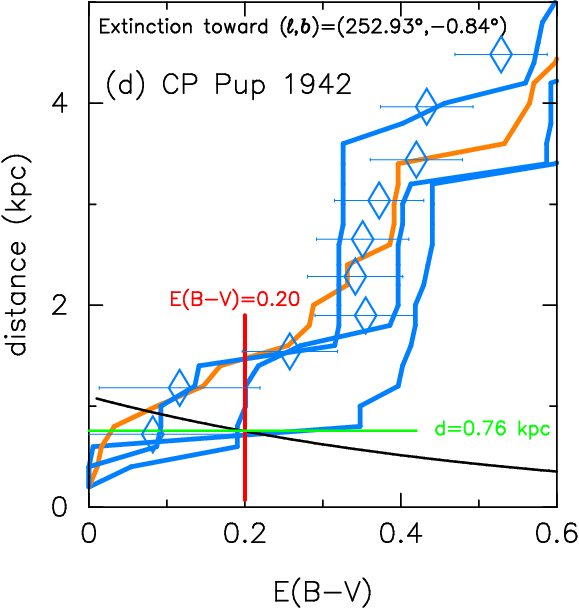}{0.41\textwidth}{}
          }
\caption{
The distance-reddening relations toward (a) V1500 Cyg, (b) V1674 Her,
(c) CP Lac, and (d) CP Pup.
The black lines denotes the relation of Equation 
(\ref{distance_modulus_extinction}) together with $(m-M)_V$ of each nova.
The thin magenta lines are
the sample distance-reddening relations given by \citet{gre19}
while the thick magenta line is their best-fit line.
We add four distance-reddening relations
(thick cyan-blue lines) of \citet{chen19hu}, which correspond to four
nearby directions toward each nova.
The unfilled cyan-blue diamonds with error bars
represent the relation of \citet{ozd18}.
(a) V1500 Cyg: the black line of $(m-M)_V=12.3$ crosses the magenta lines
at $d=1.56$ kpc and $E(B-V)=0.43$.
The filled red circles denote the distance
and reddening of nearby stars given by \citet{you76}.
(b) V1674 Her: the black line of $(m-M)_V=16.3$
crosses the magenta lines at $d=8.9$ kpc and $E(B-V)=0.5$. 
(c) CP Lac: the black line of $(m-M)_V=11.0$
crosses the magenta lines at $d=1.1$ kpc and $E(B-V)=0.24$.
See the main text for details.
(d) CP Pup: the black line of $(m-M)_V=10.0$
broadly crosses the cyan-blue lines at $d=0.76$ kpc and $E(B-V)=0.20$.
The thick orange line is the closest line to the galactic
coordinates of CP Pup $(\ell, b)=(252 \fdg 926, -0 \fdg 835)$
among the four cyan-blue lines.
\label{distance_reddening_v1500_cyg_v1674_her_cp_lac_cp_pup_v}}
\end{figure*}

In Figure \ref{distance_reddening_v1500_cyg_v1674_her_cp_lac_cp_pup_v}(a),
the distance-reddening relation (black line) for $(m-M)_V=12.3$
crosses \citet{gre19}'s Galactic 3D extinction map line (thick magenta line)
toward V1500 Cyg at the distance of $d=1.56$ kpc and the reddening of
$E(B-V)=0.43$.
This distance is consistent with the Gaia DR3 geometric distance
$d=1.57_{-0.19}^{+0.27}$ kpc \citep{bai21rf}, as listed in Table
\ref{table_brightness_novae},  and the
extinction is very close to the value of $E(B-V)=0.45$ obtained
by \citet{tom76wl} from the equivalent width of interstellar \ion{K}{1} line.

The absolute peak $V$ magnitude of V1500 Cyg is calculated to be
$M_{V, \rm max}=m_{V,\rm max} - (m-M)_{V,\rm V1500~Cyg} = 1.9 - 12.3 = -10.4$. 
Here, we adopt the maximum apparent $V$ magnitude of $m_{V,\rm max}=1.9$
from Table 5 of \citet{ozd18}. 
We plot the position of V1500 Cyg \citep[taken from ][]{del20i} in
Figure \ref{max_t2_selvelli2019_schaefer2018_hachisu_plus_saio_2fig},
which is outside \citet{hac20skhs}'s MMRD region in Figure
\ref{max_t2_selvelli2019_schaefer2018_hachisu_plus_saio_2fig}(a)
and above the blue line in Figure 
\ref{max_t2_selvelli2019_schaefer2018_hachisu_plus_saio_2fig}(b).
Thus, we confirm that V1500 Cyg is a superbright nova.

The superbrightness originates from an optically thick shocked shell.
The shell photosphere emits photons like in a hydrogen recombination
front of a Type II Plateau supernova (SN IIP) \citep[e.g.,][]{dub25}.
The spectrum is approximated by a blackbody at the temperature of
$T_{\rm BB}\sim$5000-10000 K \citep{hac26kv1674her3}.
The shocked shell becomes optically thin on day 4-5 \citep{hac26kv1674her3}.
The spectrum is replaced with that of free-free emission from the nova winds
again.  This is the reason
why \citet{gal76} and \citet{enn77} detected the transition from
the blackbody to free-free emission on day $\sim$4-5.  The main optical
emission mechanism at the optical maximum is photospheric blackbody emission
and not free-free emission in a superbright nova. 
Figure \ref{max_t2_selvelli2019_schaefer2018_hachisu_plus_saio_2fig}(a)
shows that the position of V1500 Cyg is outside
\citet{hac20skhs}'s MMRD region calculated based on the free-free emission
model light curves.  Thus, 
Figure \ref{max_t2_selvelli2019_schaefer2018_hachisu_plus_saio_2fig}(a) 
is a useful tool to identify the main emission mechanism at optical maximum.


\section{V1674 Her 2021}
\label{full_v1674_her_2021}



The 2021 outburst of V1674 Her was discovered at 8.4 mag on UT 2021
June 12.537 (=HJD 2459378.037) by Seiji Ueda (cf. CBET No.4976), which
is 0.367 days after the outburst \citep[the day zero is 
$t=0=t_{\rm OB}=$HJD 2459377.68; ][]{kat25hsa}.
It has been observed from radio, optical, UV, and X-ray, to gamma-ray
\citep{dra21, woo21, lin22, pat22, ori22, sok23, bha24, hab24, qui24}.
One of the remarkable features of V1674 Her is rich observational data
in the very early pre-discovery phase of the outburst (day 0.016--0.25),
as shown in Figure
\ref{v1674_her_v838_her_v597_pup_v5589_sgr_full_model_fit_linear}(a).
The $V$ (green squares) and visual (black stars) data are
taken from the archive of 
the American Association of Variable Star Observers (AAVSO).
Evryscope $g$ (blue filled circles)
data are from \citet{qui24}.
Here, we obtained the absolute $V$ magnitude with the distance modulus
in the $V$ band of $(m-M)_V=16.3$ \citep{kat25hsa}.

Figure \ref{v1674_her_v838_her_v597_pup_v5589_sgr_full_model_fit_linear}(a)
shows 1.35 $M_\sun$ WD light curve models 
with the three mass-accretion rates of
$1\times 10^{-11}$ (thick gray line),
$5\times 10^{-10}$ (orange line), and
$5\times 10^{-9}$ (black line) $M_\sun$ yr$^{-1}$ \citep{kat26hsa}.
These light curves are calculated with Equation (\ref{free-free_flux_v-band})
based on the free-free emission light curve model 
and the lowest $\dot{M}_{\rm acc}$ model reproduces well the rising and 
decay phases, but does not reach the optical peak.
\citet{hac26kv1674her3} showed that a strong shock arises
0.32 days after the outburst and the shocked shell becomes optically thick
soon after the shock arises.  They calculated the $V$ light curve
of an optically thick shocked shell, which is depicted by the red line
labeled shock in Figure 
\ref{v1674_her_v838_her_v597_pup_v5589_sgr_full_model_fit_linear}(a).
The shell becomes optically thin after day $\sim$1.0 and free-free emission
dominates again the $V$ band luminosity.  In Figure 
\ref{v1500cyg_1674her_model_v_observation_logscale}(b), we also plot this
free-free emission model light curve (black line labeled ``free-free
emission'') of a 1.35 $M_\sun$ WD
with $\dot{M}_{\rm acc}=1\times 10^{-11} ~M_\sun$ yr$^{-1}$ and the optically
thick shocked shell model (red line labeled shock). 

\citet{hac26kv1674her3} argued that the evolution of
spectral features on day 0.4--1.0 of V1674 Her \citep{mun21vd,ayd21sc,hab24}
supports the transition of a shocked shell from optically thick to thin
on day $\sim 1.0$.  
Also, this transition from optically thick to thin is
essentially the same as that of V1500 Cyg described in Section 
\ref{full_v1500_cyg_1975}.

We confirm again that the best fit distance modulus in the
$V$ band is $(m-M)_V=16.3\pm 0.2$ when we change $(m-M)_V$ by a 0.1 mag
step from $(m-M)_V=15.3$ to $(m-M)_V=17.3$ and, in each step, we directly
fit our nova model light curve (thick gray line and red line) with the
observation in Figure 
\ref{v1674_her_v838_her_v597_pup_v5589_sgr_full_model_fit_linear}(a).
We also obtain a similar resut of $(m-M)_V=16.3\pm 0.2$
by the time-stretching method in Appendix \ref{distance_v1674_her}.

We calculate the absolute peak $V$ magnitude to be $M_{V, \rm max}=
m_{V,\rm max} - (m-M)_{V,\rm V1674~Her} = 6.1 - 16.3 = -10.2$. 
Here, we adopt the maximum apparent $V$ magnitude of $m_{V,\rm max}=6.1$
from CBET No.4977.  We plot the position of V1674 Her in Figure 
\ref{max_t2_selvelli2019_schaefer2018_hachisu_plus_saio_2fig},
which is outside of the MMRD region \citep{hac20skhs} in Figure
\ref{max_t2_selvelli2019_schaefer2018_hachisu_plus_saio_2fig}(a). 
Also it is above the blue line in Figure 
\ref{max_t2_selvelli2019_schaefer2018_hachisu_plus_saio_2fig}(b).
Thus, we confirm that V1674 Her is a superbright nova.

%
%

With the distance modulus $(m-M)_V=16.3$ determined by the above direct fit
of our model $V$ light curve with the observation, 
we obtain the distance and reddening toward V1674 Her to be
$d=8.9$ kpc and $E(B-V)=0.50$ from the cross point between the black line
(Equation (\ref{distance_modulus_extinction})) 
and \citet{gre19}'s extinction map (magenta lines) in Figure 
\ref{distance_reddening_v1500_cyg_v1674_her_cp_lac_cp_pup_v}(b).
This extinction is close to the value of $E(B-V)=0.55$ obtained
by \citet{mun21vd} from the equivalent width of interstellar \ion{K}{1} line.
Also the reddening of $E(B-V)= 0.5$ is supported by \citet{schlaf11f}'s
Galactic 2D reddening map of $E(B-V)= 0.4985\pm 0.0191$ toward V1674 Her.

We check the distance with the Gaia parallax.
The Gaia negative parallax ($\varpi=(-0 \farcs 93628335\pm 
0 \farcs 6273195)\times 10^{-3}$) of ID 4514092717838547584 star
($m_{\rm G}=$19.95) gives a geometric distance of $d=$6000 (3242-9802) pc
\citep{bai21rf}. We list the Gaia geometric distances of each nova 
in Table \ref{table_brightness_novae}.
It should be noted again that the Gaia parallax gives 
a reasonable/correct distance if the parallax error is relatively small,
say $<$30\% \citep{schaefer22}. 
However, if the parallax is negative or its error is large, its distance
estimate depends largely on the prior information of stars (Galactic
star distribution) in the Bayesian statistics calculation \citep{bai21rf}.  
This is the reason why a negative parallax gives a positive distance. 
See \citet{schaefer18, schaefer22} for a critical explanation of
the Gaia distance.

\section{CP Lac 1936}
\label{full_cp_lac_1936}


The 1936 outburst of CP Lac was discovered 
first by K. Gomi (IAUC No.603), when it rose
to 4th mag on UT 1936 June 18.486 ($=$ JD 2428337.986).
This nova further rose to 1.9 mag on UT 1936 June 20.542 
($=$ JD 2428340.042, IAUC No.603).  In the present paper, we assume
the outburst day of CP Lac to be $t_{\rm OB}=$ JD 2428337.0.
We plot the visual (green circles) and photoelectric (filled magenta stars) 
magnitudes 
in Figure
\ref{v1500_cyg_cp_lac_cp_pup_v5583_sgr_full_model_fit_linear}(b).
We assume the distance modulus in the $V$ band of $(m-M)_V=11.0$. 
The visual data are taken from AAVSO while the photoelectric data
are from \citet{art63k}.  The decline is smooth and rapid.
\citet{mcl50} listed this nova as a prototype of the speed class
of fast novae with smooth decline.

We overplot the 1.25 $M_\sun$ WD models with the five mass-accretion rates
in Figure \ref{v1500_cyg_cp_lac_cp_pup_v5583_sgr_full_model_fit_linear}(b).
We try to fit the model $V$ light curve with the observation
by shifting the zero time point ($t=0$)
of each model horizontally back and forth.
Among the five model $V$ light curves, we adopt the blue line
of $\dot{M}_{\rm acc}=1\times 10^{-9} ~M_\sun$ yr$^{-1}$.
The optical peak is reproduced with the free-free emission
model light curve (blue line).   In other words, the good 
fitting to the free-free light curve is an evidence
that the shocked shell is optically thin near/around the optical maximum 
and do not contribute to the peak luminosity.  
We have examined the best fit distance modulus in the $V$ band
to be $(m-M)_V=11.0\pm 0.3$ when we change $(m-M)_V$ by a 0.1 mag step from
$(m-M)_V=10.0$ to $(m-M)_V=12.0$ and, in each step, we fit our model
light curve (blue line) with the observation.

Then, we calculate the absolute peak $V$ magnitude to be $M_{V, \rm max}=
m_{V,\rm max} - (m-M)_{V,\rm CP~Lac} = 1.9 - 11.0 = -9.1$,
where we adopt $m_{V,\rm max}= 1.9$ from IAUC No.603.  
In Figure \ref{max_t2_selvelli2019_schaefer2018_hachisu_plus_saio_2fig},
we have already plotted the two positions of CP Lac from the results by
\citet{schaefer18} and \citet{sel19}, which show slightly different
$t_2$ and $M_{V,\rm max}$ values from our results, where 
CP Lac is denoted by red symbols encircled by yellow lines.
However, these two are both inside \citet{hac20skhs}'s MMRD region in Figure
\ref{max_t2_selvelli2019_schaefer2018_hachisu_plus_saio_2fig}(a)
and below the blue line in Figure 
\ref{max_t2_selvelli2019_schaefer2018_hachisu_plus_saio_2fig}(b).
We confirm that CP Lac is not a superbright nova but a normal nova.

We examine the best-fit mass accretion rate to the WD of 
$\dot{M}_{\rm acc}=1\times 10^{-9} ~M_\sun$ yr$^{-1}$ in CP Lac.
\citet{honey98} found that the brightness of CP Lac in quiescence
declined by 1.2 mag and lasted for about 450 days.
They identified this object as a nova-like VY Sculptoris type
cataclysmic variable star because of the slow ingress.
From a photometry, \citet{rod05t}
suspected the presence of a 0.127 days periodicity 
together with a slight dip in the phase folded light curve.
This photometric period is, however, not orbital since radial velocity
measurements of \citet{peters06} yielded a period of 0.145143(1) days. 
These natures of CP Lac remind us of the classical nova YZ Ret
that belongs to the VY Scl type and has a similar
orbital period of $P_{\rm orb}= 0.1324539\pm 0.0000098$ days 
\citep[$= 3.179$ hr, ][]{schaefer22}.
The mass-accretion rate onto the WD in YZ Ret is estimated to be
${\dot M}_{\rm acc} \gtrsim 2 \times 10^{-9} ~M_\sun$ yr$^{-1}$
\citep{kat22shapjl}, because dwarf nova outbursts are suppressed
in a nova-like VY Scl star \citep[e.g.,][]{osa96}.
\citet{hac23k} estimated the WD mass of YZ Ret
to be $M_{\rm WD}\sim 1.33 ~M_\sun$.  On the other hand, the WD mass
of CP Lac is close to $M_{\rm WD}=1.25 ~M_\sun$.  Considering the 
similarities in the WD mass and orbital period between these two objects,
the mass accretion rate of $\dot{M}_{\rm acc}\sim 1\times 10^{-9} ~M_\sun$
yr$^{-1}$ in CP Lac is broadly consistent with that in YZ Ret, both of which
are constrained by the VY Scl type nature.

Adopting the distance of $d= 1.13\pm 0.03$ kpc from the Gaia DR3 parallax
\citep{bai21rf}, we obtain the reddening toward CP Lac
to be $E(B-V)=0.24$ from the cross point between $d=1.1$ kpc (horizontal
green line labeled ``d=1.1 kpc'') and \citet{gre19}'s extinction map
(magenta lines) in Figure 
\ref{distance_reddening_v1500_cyg_v1674_her_cp_lac_cp_pup_v}(c).
The extinction is consistent with the value of $E(B-V)=0.28\pm 0.06$
obtained by \citet{sel13} from a mean of previous estimates of reddening.
Substituting $d=1.1$ kpc and $E(B-V)=0.24$ into Equation
(\ref{distance_modulus_extinction}), we obtain $(m-M)_V=11.0$, being
consistent with our direct $V$ light curve fit result mentioned above.


%
%
%

\section{CP Pup 1942}
\label{full_cp_pup_1942}

CP Pup was discovered by B. H. Dawson at 0.5 mag on UT 1942 November 9
($=$ JD 2430672.5). It was already 1.1 mag the previous night (IAUC No.925).  
The outburst day is not well constrained.  In the present paper,
we assume the outburst day of $t_{\rm OB}=$
JD 2430671.0 ($=$ UT 1942 November 7.5).
We plot the visual light curve of CP Pup in Figure
\ref{v1500_cyg_cp_lac_cp_pup_v5583_sgr_full_model_fit_linear}(c).
It seems that there are two optical peaks, that is,
the first peak is 0.5 mag on day $\sim$2 (JD 2430673.0),
and the second peak is 0.4 mag on day $\sim$5 (JD 2430676.0),
separated by 3 days.  
This kind of two peaks in the $V$ light curves
are also seen in the classical nova V339 Del 2013 \citep{mun13hd, mun15mm, 
sko14dt, hac24km}, of which the two peaks are separated by 2 days.

Assuming the distance modulus in the $V$ band of $(m-M)_V=10.0$,
we fit the five model light curves of a 1.25 $M_\sun$ WD 
in Figure \ref{v1500_cyg_cp_lac_cp_pup_v5583_sgr_full_model_fit_linear}(c).
The thin orange line of $\dot{M}_{\rm acc}=5\times 10^{-11} ~M_\sun$ yr$^{-1}$ 
reproduces the first peak
but do not the second peak on day 5.
We then horizontally shift the thin orange line by 3 days, which is shown 
by the thick orange line.  We are able to broadly fit this thick orange 
line with the second peak from day 4 to day 9,
but the decay in the visual magnitude
is too fast after day 9 (in the visual light curve of the magenta stars).  
Thus, we regard that the first peak is the main peak of the outburst
and the second peak is a kind of additional outburst that lasts for 
$\sim$5 days.  The model $V$ light curve (thin orange line) begins to
again follow the visual observation after day 10.

We have examined the best fit distance modulus in the $V$ band
to be $(m-M)_V=10.0\pm 0.3$ when we change it by a 0.1 mag step from
$(m-M)_V=9.0$ to $(m-M)_V=11.0$ and, in each step, we fit our model
light curve (thin orange line and thick orange line) with the observation.
The visual light curve of CP Pup is reproduced only with the free-free
emission model light curve, which suggests that the shocked shell
is optically thin near/around the optical peak.
Then, we calculate the absolute peak $V$ magnitude to be $M_{V, \rm max}=
m_{V,\rm max} - (m-M)_{V,\rm CP~Pup} = 0.4 - 10.0 = -9.6$. 
Here, we adopt the maximum apparent $V$ magnitude of 
$m_{V,\rm max}=0.4$ of CP Pup from Table 1 of \citet{sel19}.
We confirm that CP Pup is not a superbright nova but a normal nova.
We list them in Table \ref{table_brightness_novae}.

%
%
%
%

In Figure \ref{v1500_cyg_cp_lac_cp_pup_v5583_sgr_full_model_fit_linear}(c),
we obtain $(m-M)_V=10.0\pm 0.3$ by the direct light curve fitting.  
The Gaia DR3 geometric distance is well determined as $d= 0.76\pm0.01$ kpc 
\citep{bai21rf}, as listed in Table \ref{table_brightness_novae}.
The crossing point between  $(m-M)_V=10.0$ (black line)
and $d= 0.76$ kpc (horizontal green line) gives the reddening of
$E(B-V)=0.20$ in Figure
\ref{distance_reddening_v1500_cyg_v1674_her_cp_lac_cp_pup_v}(d).
This value is very consistent with $E(B-V)=0.20\pm 0.04$ obtained
by \citet{sel13} from a mean of previous estimates of reddening.
Therefore, we adopt the reddening of $E(B-V)=0.20$ for CP Pup.

If we use $E(B-V)=0.10$ listed by \citet{schaefer22},
we obtain $(m-M)_V=9.7$ mag
and the absolute peak $V$ magnitude of $M_{V, \rm max}=
m_{V,\rm max} - (m-M)_{V,\rm CP~Pup} = 0.4 - 9.7 = -9.3$. 
In Figure \ref{max_t2_selvelli2019_schaefer2018_hachisu_plus_saio_2fig},
we have already plotted the two positions of CP Pup from the results by
\citet{schaefer18} and \citet{sel19}, which show slightly different
$t_2$ and $M_{V,\rm max}$ values from our results, where
CP Pup is denoted by red symbols encircled by cyan lines.
These two positions are both inside \citet{hac20skhs}'s MMRD region in Figure
\ref{max_t2_selvelli2019_schaefer2018_hachisu_plus_saio_2fig}(a)
and below the blue line in Figure 
\ref{max_t2_selvelli2019_schaefer2018_hachisu_plus_saio_2fig}(b).

Our best fit mass accretion rate to the WD is $\dot{M}_{\rm acc}=
5\times 10^{-11}$ $M_\sun$ yr$^{-1}$ (orange line) in Figure 
\ref{v1500_cyg_cp_lac_cp_pup_v5583_sgr_full_model_fit_linear}(c). 
Recently, \citet{ver24sh} obtained 
$\dot{M}_{\rm acc}=($1--2$)\times 10^{-10}$ $M_\sun$ yr$^{-1}$
from their updated X-ray analyses, in which
they adopted $M_{\rm WD}=0.78 ~M_\sun$ and $R_{\rm WD}=0.011 ~R_\sun$.
See also their Table 4 for a summary of the other estimates
on the mass-accretion rates.
If we convert their value with our values of
$M_{\rm WD}=1.25 ~M_\sun$ and $R_{\rm WD}=0.0049 ~R_\sun$
\citep[$\log R_{\rm WD}/R_\sun = -2.312$ in Table 1 of ][]{hac26kmaxej}.
we obtain $\dot{M}_{\rm acc}=$(3--6)$\times 10^{-11}$ $M_\sun$ yr$^{-1}$
because $\dot{M}_{\rm acc}$ is reduced by a factor of 3.6
($=(1.25/0.78)/(0.0049/0.011))$ from $L_{\rm X}\propto G M_{\rm WD}
\dot{M}_{\rm acc}/ R_{\rm WD}$.
This value is very consistent with our light curve fitting.
We summarize our results in Table \ref{table_brightness_novae}.


\begin{figure*}
\gridline{\fig{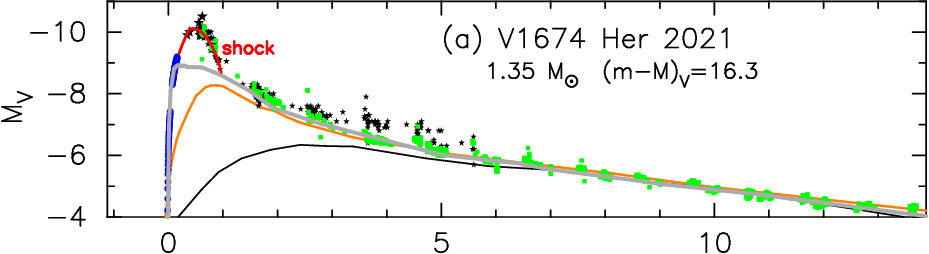}{0.75\textwidth}{}
          }
\gridline{
          \fig{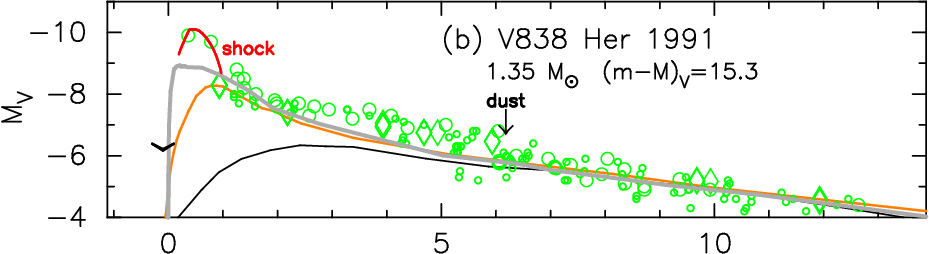}{0.75\textwidth}{}
          }
\gridline{
          \fig{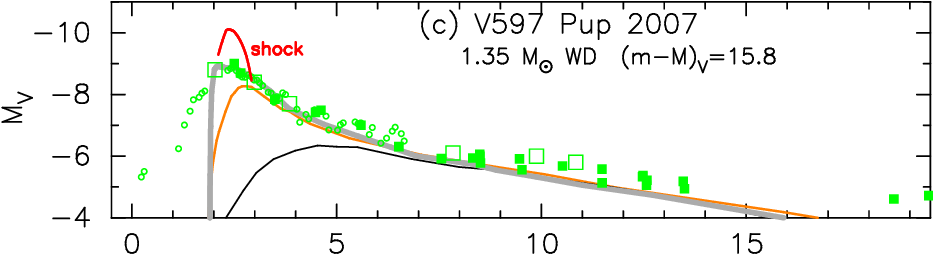}{0.75\textwidth}{}
          }
\gridline{
          \fig{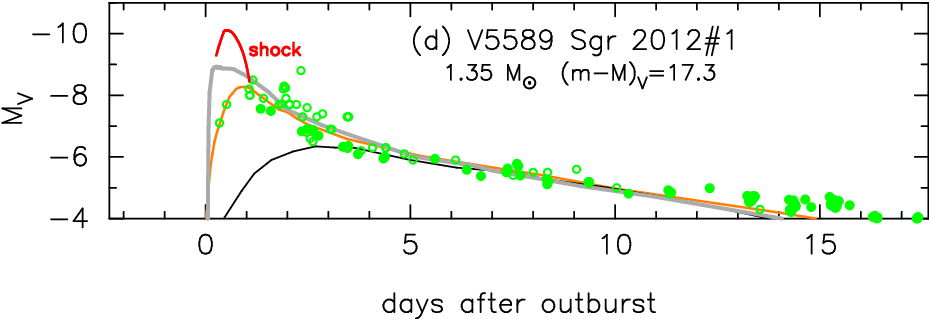}{0.75\textwidth}{}
          }
\caption{
The $V$/$g$/visual/SMEI light curves of (a) V1674 Her, (b) V838 Her,
(c) V597 Pup, and (d) V5589 Sgr on a linear timescale. 
We add three model $V$ light curves of a 1.35 $M_\sun$ WD with
mass-accretion rates of $1\times 10^{-11}$ (thick gray line),
$5\times 10^{-10}$ (orange line), and $5\times 10^{-9}$ (black line)
$M_\sun$ yr$^{-1}$.  The thick red line labeled ``shock'' indicates
the $V$ brightness of the optically thick shocked shell in V1674 Her
\citep{hac26kv1674her3}.
(a) V1674 Her: the $V$/$g$/visual data are the same as those in Figure
\ref{v1500cyg_1674her_model_v_observation_logscale}(b).
(b) V838 Her: the data are the same as those
in Figure 8 of \citet{kat09v838her}.  The downward black caret symbol
shows the upper limit of $m_v > 9$ on day $-0.1$ \citep{ueta91}.
(c) V597 Pup: the optical data are the same as those in Figure 20
of \citet{hac10k}.
(d) V5589 Sgr: the $V$ (filled green circles) and visual (open green circles)
light curves.  See the main text for the sources of each optical data.
\label{v1674_her_v838_her_v597_pup_v5589_sgr_full_model_fit_linear}}
\end{figure*}

\section{V838 Her 1991}
\label{full_v838_her_1991}


V838~Her is a very fast nova with $t_2=1$~day and $t_3=4$~days,
and its peak brightness reached $m_{V, \rm max}=5.3$ \citep[e.g.,][]{str10}.
The orbital period was obtained to be $P_{\rm orb}=0.2976$~days
\citep[$=7.14$ hr, ][]{ing92, lei92}.

The 1991 outburst of V838 Her was discovered independently
by \citet{sug91} on UT 1991 March 24.78 (=JD 2448340.28) at 5.4 mag
and by \citet{alc91} at $\sim 5$ mag on UT 1991 March 25.67.
G. Alcock's visual discovery around the peak was made in strong twilight,
so the peak magnitude was not accurate. The outburst time can be estimated
from the upper limit (9 mag) prediscovery observation by
\citet{ueta91} on JD 2448339.9 and M. Sugano's discovery on
JD 2448340.281. In the present work, we adopt the outburst
time of $t_{\rm OB}= (t=0=)$JD 2448340.0 as day zero.

Figure \ref{v1674_her_v838_her_v597_pup_v5589_sgr_full_model_fit_linear}(b)
shows the $V$ and visual light curves of V838 Her on a linear timescale
with the distance modulus in the $V$ band of $(m-M)_V=15.3$.
We overplot the free-free emission model light curves for a 1.35 $M_\sun$ WD
with the three mass-accretion rates.  
The red line labeled shock shows the $V$ light curve for 
an optically thick shocked shell model in V1674 Her \citep{hac26kv1674her3}.
Among the three models,
the gray line of $\dot{M}_{\rm acc}=1\times 10^{-11} ~M_\sun$ yr$^{-1}$ 
together with the red line of optically thick shocked shell in V1674 Her
is close to the observed light curve of V838 Her.  In this light curve
fitting, the earliest two optical data (green circles) clearly indicate
formation of an optically thick shocked shell.  This supports 
\citet{hac26kv1674her3}'s conclusion that an optically thick shocked shell
makes a nova superbright.
It seems that there is a slight drop on day 6-7 and thereafter in the visual
and $V$ light curves of V838 Her. 
This is due to a dust shell formation as the IR observations indicate 
\citep{chand92, harri94,  kidg93, wood92, lynch92, smith95, kat09v838her}

We have examined the best fit distance modulus in the $V$ band
to be $(m-M)_V=15.3\pm 0.2$ when we change it by a 0.1 mag step from
$(m-M)_V=14.3$ to $(m-M)_V=16.3$ and, in each step, we fit our model
light curves (thick gray line and red line) with the observation.
We also check this $(m-M)_V=15.3\pm 0.2$ by the time-stretching method
in Appendix \ref{distance_v838_her}.
Then, we have $M_{V,\rm max}= 5.3 -15.3= -10.0$
from $m_{V,\rm max}= 5.3$ \citep[Table 6 of][]{ozd18}.
These results are summarized in Table \ref{table_brightness_novae}.
We also plot the position of V838 Her in Figure
\ref{max_t2_selvelli2019_schaefer2018_hachisu_plus_saio_2fig}.
V838 Her is located outside of \citet{hac20skhs}'s MMRD region in Figure 
\ref{max_t2_selvelli2019_schaefer2018_hachisu_plus_saio_2fig}(a).
This suggests that the optical peak of V838 Her cannot be
reproduced by free-free emission, but an optically thick shocked shell.
The position is also above the blue line
in Figure \ref{max_t2_selvelli2019_schaefer2018_hachisu_plus_saio_2fig}(b). 
Thus, V838 Her is a superbright nova.

%
%

\begin{figure*}
\gridline{\fig{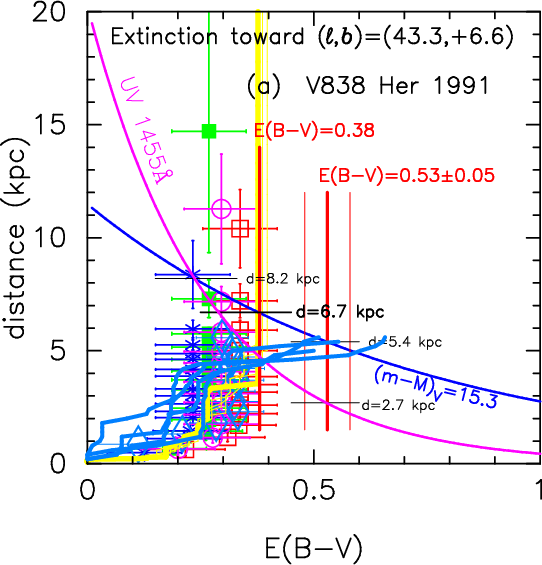}{0.4\textwidth}{}
          \fig{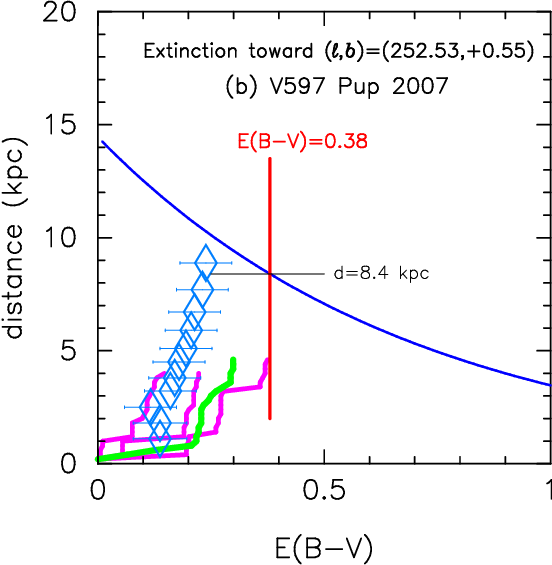}{0.4\textwidth}{}
          }
\gridline{
          \fig{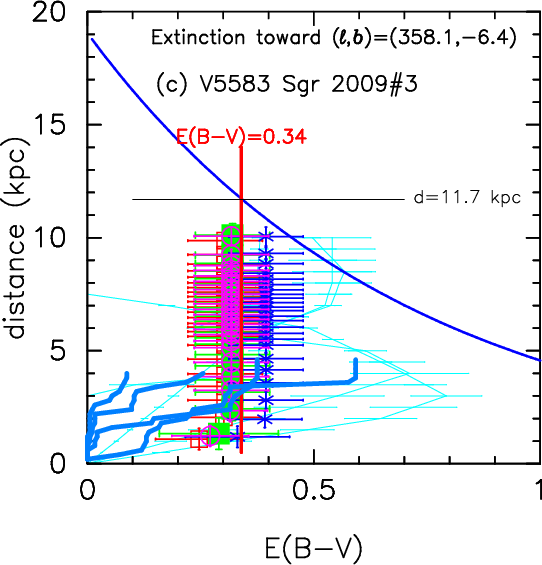}{0.4\textwidth}{}
          \fig{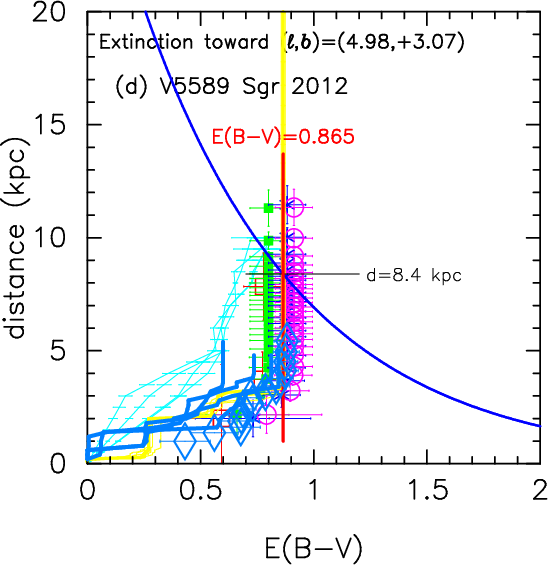}{0.4\textwidth}{}
          }
\caption{
Same as Figure \ref{distance_reddening_v1500_cyg_v1674_her_cp_lac_cp_pup_v},
but for (a) V838 Her, (b) V597 Pup, (c) V5583 Sgr, and (d) V5589 Sgr.
Thin and thick yellow lines show the distance-reddening relations of
\citet{gre19}.  Open red squares, filled green
squares, blue asterisks, and open magenta circles with error bars
represent the relations of \citet{mar06} in the four directions
close to each nova.  Solid blue line indicates the distance-reddening
relation of Equation (\ref{distance_modulus_extinction}) with each $(m-M)_V$
of a nova.  We add four distance-reddening relations (thick cyan-blue,
green, or magenta lines) of \citet{chen19hu}, which correspond to four
nearby directions toward each nova.  Open cyan-blue diamonds denote
the distance-reddening relation taken from \citet{ozd16}.
We plot four \citet{schu14}'s distance-reddening relations
toward a nova by cyan lines with error bars. 
(a) V838 Her: the thick magenta line
denotes the distance-reddening relation given by the 1455~\AA\
light curve fitting together with Equation
(\ref{uv1455_distance_modulus}), which is taken from \citet{hac18kb}.
(b) V597 Pup: the green line denotes \citet{chen19hu}'s 3D Galactic
extinction map closest toward the galactic coordinates of V597 Pup,
i.e., $(\ell,b)=(252\fdg 55, +0\fdg 55)$.  The reddening of $E(B-V)=0.38$
is the result of \citet{schlaf11f}'s 2D Galactic extinction map toward 
V597 Pup.
(c) V5583 Sgr: the reddening of $E(B-V)=0.34$
is the result of \citet{schlaf11f}'s 2D Galactic extinction map toward 
V5583 Sgr.
(d) V5589 Sgr: the blue line (Equation (\ref{distance_modulus_extinction})
with $(m-M)_V=17.3$)
crosses the yellow lines \citep{gre19} at $d=8.4$ kpc and $E(B-V)=0.865$.
\label{distance_reddening_v838_her_v597_pup_v5583_sgr_v5589_sgr_v}}
\end{figure*}

We plot the distance-reddening relation (blue line) of Equation 
(\ref{distance_modulus_extinction}) with $(m-M)_V=15.3$ in Figure
\ref{distance_reddening_v838_her_v597_pup_v5583_sgr_v5589_sgr_v}(a).
This blue line crosses \citet{gre19}'s relations (yellow lines) 
of 3D Galactic extinction map toward V838 Her at $d=6.7$ kpc and
$E(B-V)=0.38$.  The reddening of $E(B-V)=0.38$ is consistent with
\citet{schlaf11f}'s Galactic 2D reddening map of $E(B-V)=0.3681\pm 0.0024$
toward V838 Her.  
The other reddenings were given by several authors
\citep[e.g.,][]{mat93, harri94, van96, kat09v838her}, which lies between
$E(B-V)=0.3$--$0.7$.
The distance estimated by the Gaia DR3 parallax of 
ID 4504548029183559552 star ($m_{\rm G}=20.07$)
$\varpi = (0\farcs 8631003 \pm 0\farcs 83557326)\times 10^{-3}$
is not accurately constrained to be 
$d_{\rm geometric}= 5013.440$ (2565.142-7146.869) pc or
$d_{\rm photogeometric}= 6241.201$ (5114.441-7706.806) pc \citep{bai21rf},
but the both distance estimates are broadly consistent
with our $d=6.7$ kpc.
We summarize these data of V838 Her in Table \ref{table_brightness_novae}.

\citet{kat09v838her} obtained a relatively larger value of
$E(B-V)=0.53\pm0.05$ from the 2175~\AA\ feature of the UV spectra of V838~Her 
on day 10, 12, 15, and 22.  They also derived
the distance of $d=2.7\pm0.5$~kpc from the UV~1455~\AA\ flux fitting and
the WD mass of $M_{\rm WD}=1.35\pm0.02 ~M_\sun$ from the model
light curve fitting.
They fitted their model UV 1455~\AA\  light curve with the IUE observation
and obtained a relation of
\begin{equation}
-2.5\log \left( {L_{\lambda}^{\rm obs} \over L_{\lambda}^{\rm mod}} \right)
= R_\lambda E(B-V) + 5\log \left( {{d} \over {10~{\rm kpc}}} \right),
\label{uv1455_distance_modulus}
\end{equation}
where $L_{\lambda}^{\rm mod}$ is the model luminosity at the distance of
$d=10$~kpc, $L_{\lambda}^{\rm obs}$ is the observed luminosity at the
wavelength $\lambda$,
the absorption is calculated from $A_\lambda=R_\lambda E(B-V)$, and
$R_\lambda=8.3$ for $\lambda=1455$~\AA\  \citep{sea79}.
For the $1.35~M_\sun$ WD model in their Figure 10(a),
$L_{1455}^{\rm obs}=5.0$ and $L_{1455}^{\rm mod}=21.0$,
in units of $10^{-12}$~erg~cm$^{-2}$~s$^{-1}$~\AA$^{-1}$,
at the upper bound of their Figure 10(a).

We plot the distance-reddening relation of this UV 1455 \AA\  luminosity
(magenta line) in Figure 
\ref{distance_reddening_v838_her_v597_pup_v5583_sgr_v5589_sgr_v}(a). 
If we substitute $E(B-V)=0.53$ into Equation
(\ref{uv1455_distance_modulus}), we obtain the distance of $d=2.7$~kpc
as shown in Figure
\ref{distance_reddening_v838_her_v597_pup_v5583_sgr_v5589_sgr_v}(a).
The difference between \citet{kat09v838her} and the present work can be
resolved as follows:
The observed flux of UV 1455\AA\  attains its peak on day 5.8, just before the
dust shell formation on day 7.  Therefore, the peak fluxes of optical $V$ or
visual bands and UV 1455\AA\  are not affected by a newly formed dust shell.
In Figure \ref{distance_reddening_v838_her_v597_pup_v5583_sgr_v5589_sgr_v}(a),
the magenta line of UV 1455\AA\  crosses the blue line of the $V$ band
at $d=8.2$ kpc and $E(B-V)=0.24$, suggesting that the reddening is rather
small, as small as 0.3.  
We suppose that the larger value of
$E(B-V)=0.53$ could be a result of a dust shell formation after day 7.
We adopt a value of $E(B-V)=0.38$, which is only the interstellar
reddening.  The excess part of $\Delta E(B-V)= 0.53 - 0.38= 0.15$ could
be due to the newly formed circumstellar dust shell (after day 7).
Then, the $V$ band absorption is estimated to be 
$\Delta A_V= 3.1 \Delta E(B-V) = 3.1\times 0.15 \approx 0.5$ mag.
The brightness drop by the optically thin dust shell formation 
is broadly consistent with the gap before/after day 7 in Figure 
\ref{v1674_her_v838_her_v597_pup_v5589_sgr_full_model_fit_linear}(b).

\section{V597 Pup 2007}
\label{full_v597_pup_2007}


The 2007 outburst of V597 Pup was discovered by \citet{per07}
at a visual magnitude of 7.0 on UT 2007 November 14.23
($=$ JD 2454417.73), reaching a peak visual magnitude of
$m_v=6.4$ on UT 2007 November 14.48 \citep[$=$ JD 2454417.98: ][]{young07}.
The nova then declined rapidly with $t_2=2.5$ days \citep{naik09}.
A pre-eruption detection is found within the Digitized Sky Survey
\citep{young07} with a source at $V\sim 20$, which coincides
with the nova position. \citet{war09} revealed the nova as an intermediate
polar of the orbital period of $P_{\rm orb}=2.67$ hr, with a spin
period of 8.7 min ($=524$ s).
It became visible in soft X-rays in January 2008 \citep{ness08}.
\citet{hou16db} presented the Solar Mass Ejection Imager (SMEI)
light curve of V597 Pup before and after the optical maximum.

We plot the $V$, visual, and SMEI light curves of V597 Pup in Figure
\ref{v1674_her_v838_her_v597_pup_v5589_sgr_full_model_fit_linear}(c)
with the distance modulus of $(m-M)_V=15.8$.
We overplot the three free-free emission model light curves of a
1.35 $M_\sun$ WD
and the optically thick shocked shell model (red line). 
Among the three models 
the thick gray line is close to the observed light curve
of V597 Pup except for the very early rising phase.
In this light curve fitting, we do not need 
the red line light curve labeled shock.
The decay of the observed light curve (green symbols)
is slightly slower than
that of the gray line, suggesting that the WD mass of V597 Pup is slightly
smaller (less massive) than 1.35 $M_\sun$.  
Thus, we may conclude that the 1.35 $M_\sun$ or slightly smaller mass WD
with $\dot{M}_{\rm acc}\sim 1\times 10^{-11} ~M_\sun$ yr$^{-1}$ 
is suitable for the $V$/visual light curve of V597 Pup.

We have examined the best fit distance modulus in the $V$ band
to be $(m-M)_V=15.8\pm 0.2$ when we change it by a 0.1 mag step from
$(m-M)_V=14.8$ to $(m-M)_V=16.8$ and, in each step, we fit our model light
curve (thick gray line) with the observation in Figure 
\ref{v1674_her_v838_her_v597_pup_v5589_sgr_full_model_fit_linear}(c).
We have checked $(m-M)_V=15.8\pm0.2$ by the time-stretching
method in Appendix \ref{distance_v597_pup}.

%
%

The absolute $V$ peak is $M_{V,\rm max}= 6.8 -15.8= -9.0$,
where $m_{V,\rm max}= 6.8$ is taken from the Variable Star Oververs 
League of Japan (VSOLJ) 
data.\footnote{\url{http://www.cetus-net.org/cgi-bin/obs_search.cgi}}
These results are summarized in Table \ref{table_brightness_novae}.
We also plot the position of V597 Pup in Figure
\ref{max_t2_selvelli2019_schaefer2018_hachisu_plus_saio_2fig}.
V597 Pup is located inside of \citet{hac20skhs}'s MMRD region
in Figure \ref{max_t2_selvelli2019_schaefer2018_hachisu_plus_saio_2fig}(a)
and below the blue line
in Figure \ref{max_t2_selvelli2019_schaefer2018_hachisu_plus_saio_2fig}(b). 
Thus, V597 Pup is not a superbright nova but a normal nova.

Figure \ref{distance_reddening_v838_her_v597_pup_v5583_sgr_v5589_sgr_v}(b)
shows various distance-reddening relations toward V597 Pup.
The 3D Galactic extinction map of \citet{gre19} is not available
toward V597 Pup.  It seems that the reddening saturates at $d > 8$ kpc.
If we use \citet{schlaf11f}'s Galactic 2D extinction map,
we have $E(B-V)=0.38$.  The cross point of our blue line, 
Equation (\ref{distance_modulus_extinction}) together with 
$(m-M)_V=15.8$, and the vertical red line of $E(B-V)=0.38$ gives
the distance of $d=8.4$ kpc.  We summarize these results in Table
\ref{table_brightness_novae}.

\citet{rud08} 
obtained spectroscopic observations spanning 0.8--2.42 $\mu m$
on UT 2008 January 7.48, and showed that the \ion{O}{1} lines 
indicate a small reddening of $E(B-V)\sim 0.3$ and there are no
indications of dust formation in the nova.  This is consistent with
the above reddening of $E(B-V)=0.38$.

The distance estimated by the Gaia DR3 parallax of 
ID 5546402675939272576 star ($m_{\rm G}=20.03$)
$\varpi = (0\farcs 30288753 \pm 0\farcs 39470413)\times 10^{-3}$
is not accurately constrained to be 
$d_{\rm geometric}= 4800.752$ (2901.494-6757.547) pc or
$d_{\rm photogeometric}= 11309.661$ (4844.784-18346.668) pc
\citep{bai21rf}.  We suppose that the Gaia parallax is not accurate
enough to reject our estimated distance of $d=8.4$ kpc because
the parallax error is larger than the parallax itself and therefore
their results depend strongly on the prior information of star 
distribution in our Galaxy \citep[see, e.g.,][]{schaefer22}.

\section{V5583 Sgr 2009\#3}
\label{full_v5583_sgr_2009_no3}


V5583 Sgr is a very fast nova, the 2009 outburst of which
was discovered on UT 2009 August 6.49
(JD 2455049.99) by \citet{nis09} at mag $\sim$7.7. 
Their survey frames on July 22.5 and 29.6 showed nothing
at this position (limiting mag 12.7), and nothing is visible on
Digitized Sky Survey images.  The ASAS3V images showed $V = 7.78$
on August 6.2 \citep{nis09}.  It then rose to maximum on 
UT August 7.57 (JD 2455051.07) at $m_V=7.43$ \citep{maehara09}.
An approximate $t_2=5$ days is given by \citet{schw11}.
It became visible in soft X-rays in UT 2009 October 27, about
80 days after maximum, and went off in UT 2010 April 30,
about 270 days after maximum \citep{schw11}.
\citet{hou16db} presented the SMEI light curve of V5583 Sgr
before and after the optical maximum.

Figure \ref{v1500_cyg_cp_lac_cp_pup_v5583_sgr_full_model_fit_linear}(d)
shows the $V$, visual, and SMEI light curve (green symbols) of V5583 Sgr 
with the distance modulus in the $V$ band of $(m-M)_V=16.4$.
We overplot the free-free emission model light curves of 
a 1.25 $M_\sun$ WD with the five mass-accretion rates. 
Among the five model $V$ light curves, we adopt the red line
of $\dot{M}_{\rm acc}=5\times 10^{-10} ~M_\sun$ yr$^{-1}$.

%
%

We have examined the best fit distance modulus in the $V$ band
to be $(m-M)_V=16.4\pm 0.1$ when we change it by a 0.1 mag step from
$(m-M)_V=15.4$ to $(m-M)_V=17.4$ and, in each step, we fit our model
light curve (red line) with the observation of V5583 Sgr in Figure
\ref{v1500_cyg_cp_lac_cp_pup_v5583_sgr_full_model_fit_linear}(d).
This result is consistent with that obtained by the time-stretching method,
$(m-M)_V=16.4\pm 0.2$, as described in Appendix \ref{distance_v5583_sgr}.

The absolute $V$ peak is $M_{V,\rm max}= 7.4-16.4= -9.0$
from $m_{V,\rm max}= 7.4$ \citep{maehara09}.
These results are summarized in Table \ref{table_brightness_novae}.
We also plot the position of V5583 Sgr in Figure
\ref{max_t2_selvelli2019_schaefer2018_hachisu_plus_saio_2fig}.
V5583 Sgr is located inside of \citet{hac20skhs}'s MMRD region in Figure 
\ref{max_t2_selvelli2019_schaefer2018_hachisu_plus_saio_2fig}(a)
and below the blue line
in Figure \ref{max_t2_selvelli2019_schaefer2018_hachisu_plus_saio_2fig}(b). 
Thus, V5583 Sgr is not a superbright nova but a normal nova.

Figure \ref{distance_reddening_v838_her_v597_pup_v5583_sgr_v5589_sgr_v}(c) 
shows various distance-reddening relations toward V5583 Sgr.
We plot 
Equation (\ref{distance_modulus_extinction}) with $(m-M)_V=16.4$.
\citet{schlaf11f} presented a Galactic 2D reddening map which gives
the reddening value of $E(B-V)=0.3371\pm 0.0049$ toward V5583 Sgr.  
The blue line of $(m-M)_V=16.4$ crosses the red line of $E(B-V)=0.34$
at the distance of $d=11.7$ kpc, where we assume that the reddening 
saturates for $d>5$ kpc.
We summarize these data of V5583 Sgr in Table \ref{table_brightness_novae}.

We check the distance of V5583 Sgr with the results of Gaia DR3 parallax,
which gives a candidate of a nearby star of source ID
4042579731909968384 with the brightness of $m_{\rm G}= 16.36$,
$1\farcs 43$ apart from the position of V5583 Sgr \citep{mro15up}.
Its parallax 
is $\varpi = (0\farcs 42908674 \pm 0\farcs 060200144)\times 10^{-3}$
and the distance is $d= 2559$ (2174--3158) pc \citep{bai21rf}.
This object ($m_{\rm G}= 16.36$) is 4 mag brighter than 
V5583 Sgr: $V=20.73$ in the pre-outburst and $V=20.31$ 
in the post-outburst \citep{mro15up}, and is unlikely to be 
the object corresponding to V5583 Sgr in quiescence.
Although \citet{schaefer22} adopted this brighter star ($m_{\rm G}= 16.36$)
as V5583 Sgr, \citet{mro15up}'s light curve of V5583 Sgr clearly shows
that the brightness of V5583 Sgr decreased gradually from 16th mag
(in UT 2010 April) to 20th mag (in UT 2014 February) in Figure 10 of 
\citet{mro15up}.  The Gaia DR3 started on UT 2014 July 25 and ended 
on UT 2017 May 28, that is, started after V5583 Sgr became fainter than
20th mag. 
We conclude that this Gaia candidate star is too bright to be compatible
with V5583 Sgr in quiescence.
Thus, we have no Gaia DR3 distance to V5583 Sgr.


\section{V5589 Sgr 2012\#1}
\label{full_v5589_sgr_2012_no1}


The 2012 outburst of V5589 Sgr was discovered by \citet{kor12s}
early on UT 2012 April 21.
Then the star reached a maximum brightness of $V\sim 8.8$
around on UT 2012 April 21.654 \citep{seach12}.
The Swift satellite made its first observations of
the nova outburst for 1500 s starting around on UT 2012 April 21.8125,
and detected ultraviolet emission at
$M2=13.90\pm0.05$, but no X-ray component \citep{sok12}.
Pre-discovery images from Xingming observatory
indicate that the outburst began as early as April 20.8403 \citep{gao12}.
V5589 Sgr reached $m_{V, \rm max}\sim9.0$
on JD~2456039.56 (UT 2012 April 22.06) \citep[e.g.,.][]{mro15up}.
\citet{wal12bt} estimated the decline rates of $t_2=4.5\pm1.5$~days
and $t_3=7$~days while \citet{wes16sc} obtained $t_2=6.8\pm0.8$~days
and $t_3=12.8\pm1.5$~days.
\citet{thomp17} analyzed Solar Terrestrial Relations Observatory
(STEREO) data and obtained optical maximum at mag $8.23\pm0.03$ in
the Heliospheric Imager (HI)-1 bandpass on JD~2456039.3224.
They also obtained $t_2 = 5.0\pm0.6$ days and $t_3 = 10.9\pm0.7$ days.
In the present paper, we adopt the outburst day of
$t_{\rm OB}=$ JD 2456038.0 (UT 2012 April 20.5) from the data of
\citet{gao12} and $t_2=4.5$ days after \citet{wal12bt},
which are listed in Table \ref{table_brightness_novae}.

V5589~Sgr was identified
as a hybrid nova from \ion{Fe}{2} to He/N type \citep{wal12bt}.
The nova had entered the coronal phase of [\ion{Fe}{10}],
[\ion{Fe}{11}],  and [\ion{Fe}{14}] by day 65 \citep{wal12bt}.
The detection of the coronal phase sometimes
indicates that the nova had already entered the supersoft X-ray
source (SSS) phase \citep[e.g.,][]{wal12bt}.
The Swift/XRT observation showed that
the nova increased its soft X-ray flux on day 64.5 \citep{wes16sc}.
We may regard that the nova had entered the SSS phase at least
by day 64.5 \citep{wal12bt}.

\citet{mro15up} obtained the orbital period to be
$P_{\rm orb}=1.5923$~days ($P_{\rm orb}=38.215$~hr).
They suggested that the companion has already evolved off
the main-sequence and is a subgiant like the recurrent nova U~Sco
\citep[$P_{\rm orb}=1.23$ day;][]{sch95r}.
\citet{mro15up} concluded that eclipses were seen,
indicating that the inclination is close to 90$\arcdeg$. They also deduced
from the eruption amplitude ($<10$ mag in the $V$ band), fast decline, long
orbital period, and broad H$\alpha$ profiles that V5589 Sgr contains a
massive WD with a high-mass transfer rate, and is a good
candidate to be a recurrent nova.

Figure \ref{v1674_her_v838_her_v597_pup_v5589_sgr_full_model_fit_linear}(d)
shows the $V$ and visual light curves (green symbols) of V5589 Sgr
with the distance modulus in the $V$ band of $(m-M)_V=17.3$.
We overplot the free-free emission model light curves of a 1.35 $M_\sun$ WD
with the three mass-accretion rates.
The red line denotes
an optically thick shocked shell for V1674 Her \citep{hac26kv1674her3}.
Among the three models,
the orange line of $\dot{M}_{\rm acc}= 5\times 10^{-10} ~M_\sun$ yr$^{-1}$
perfectly fits with the observed light curve of V5589 Sgr.


We have examined the best fit distance modulus in the $V$ band
to be $(m-M)_V=17.3\pm 0.2$ when we change it by a 0.1 mag step from
$(m-M)_V=16.3$ to $(m-M)_V=18.3$ and, in each step, we fit our model light
curve (orange line) with the observation in Figure
\ref{v1674_her_v838_her_v597_pup_v5589_sgr_full_model_fit_linear}(d).
We also check this $(m-M)_V=17.3\pm 0.2$ by the time-stretching method
in Appendix \ref{distance_v5589_sgr}.

%
%

The absolute $V$ peak is $M_{V,\rm max}= 8.8-17.3= -8.5$,
where $m_{V,\rm max}= 8.8$ is taken from CBET No.3089.
These results are summarized in Table \ref{table_brightness_novae}.
We also plot the position of V5589 Sgr in Figure
\ref{max_t2_selvelli2019_schaefer2018_hachisu_plus_saio_2fig}.
V5589 Sgr is located inside of \citet{hac20skhs}'s MMRD region in Figure 
\ref{max_t2_selvelli2019_schaefer2018_hachisu_plus_saio_2fig}(a)
and below the blue line
in Figure \ref{max_t2_selvelli2019_schaefer2018_hachisu_plus_saio_2fig}(b). 
Thus, V5589 Sgr is not a superbright nova but a normal nova.



For the reddening toward V5589~Sgr, $(l,b)=(4\fdg9766, +3\fdg0724)$,
the NASA/IPAC galactic 2D dust absorption map gives $E(B-V)=0.84\pm0.04$
\citep{schlaf11f}.
The VVV survey catalogue gives $E(B-V)= A_{K_s}/0.36= 0.274/0.36= 0.76$
\citep{sai13}.  \citet{wes16sc} estimated the reddening of
$E(B-V)= 0.8\pm 0.19$ using four diffuse interstellar band (DIB) features.

We further check the distance and reddening based on various
distance-reddening relations in Figure
\ref{distance_reddening_v838_her_v597_pup_v5583_sgr_v5589_sgr_v}(d). 
\citet{mar06}'s relations are plotted
in four directions close to the direction of V5589~Sgr:
$(l, b)=(4\fdg75,  +3\fdg00)$, $(5\fdg00,+3\fdg00)$,
$(4\fdg75,+3\fdg25)$, and $(5\fdg00,+3\fdg25)$,
which are plotted
by open red squares, filled green squares, blue asterisks,
and open magenta circles, respectively, each with error bars.
The closest direction is that of filled green squares.
We plot the 3D reddening map of \citet{schu14}
by the very thin cyan lines (four close directions toward V5589~Sgr).
\citet{gre19}'s relations are denoted by the thin solid yellow
lines, while the thick yellow line corresponds to their best fit line.
The open cyan-blue diamonds are the relation of \citet{ozd18}.
The solid cyan-blue line corresponds to the relation of \citet{chen19hu}.

We plot the distance-reddening relation (blue line) of Equation 
(\ref{distance_modulus_extinction}) with $(m-M)_V=17.3$ in Figure
\ref{distance_reddening_v838_her_v597_pup_v5583_sgr_v5589_sgr_v}(d).
This blue line crosses \citet{gre19}'s relations (yellow lines) 
of 3D Galactic extinction map toward V5589 Sgr at $d=8.4$ kpc and
$E(B-V)=0.865$.  
We summarize these data of V5589 Sgr in Table \ref{table_brightness_novae}.

The Gaia DR3 parallax of star ID 4068844517774588544 
($0\farcs 05$ apart from the position of V5589 Sgr
and the brightness of $m_{\rm G}=17.86$) gives 
$\varpi = (0\farcs 140766 \pm 0\farcs 21553609) \times 10^{-3}$.
Its distance is estimated to be $d_{\rm geometric}= 6184$ (4085--8363) pc
\citep{bai21rf}.
The parallax error is much larger than the parallax itself
and therefore its distance cannot be
accurately constrained only with the Gaia DR3 parallax as already 
explained in Section \ref{full_v1674_her_2021}.
Our obtained values of $E(B-V)=0.865$ and $d=8.4$ kpc are broadly 
consistent with \citet{wes16sc}'s $E(B-V)=0.8\pm 0.19$ and
\citet{bai21rf}'s $d=6.2_{-2.1}^{+2.2}$ kpc, respectively.



\section{Conclusions}
\label{sec_conclusion}

We have reevaluated the distance moduli in the $V$ band of six 
candidate novae, CP Lac, CP Pup, V838 Her, V597 Pup, 
V5583 Sgr, and V5589 Sgr, as listed in Table \ref{table_brightness_novae},
and found that V838 Her is a superbright nova defined by \citet{del91}.
To clearly identify the superbrightness, we have posed another requirement
that they are located outside of \citet{hac20skhs}'s theoretical region of
free-free emission nova light curves in the MMRD diagram of Figure
\ref{max_t2_selvelli2019_schaefer2018_hachisu_plus_saio_2fig}(a).
There are now three identified superbright novae in our Galaxy, i.e.,
V1500 Cyg, V838 Her, and V1674 Her, as shown in Figure
\ref{max_t2_selvelli2019_schaefer2018_hachisu_plus_saio_2fig}. 
This supports our working hypothesis that the origin of superbrightness
may be the formation of an optically thick shocked shell
around the optical peak of a nova.

\begin{acknowledgments}
We are grateful to the anonymous referee for useful comments
that improved the manuscript.  
We acknowledge with thanks the variable star observations
from the AAVSO International Database contributed by
observers worldwide and used in this research.
We also thank the VSOLJ
for the nova data used in this research.
\end{acknowledgments}

\vspace{5mm}
\facilities{Swift(XRT), AAVSO}


\appendix

\section{Time-stretching Method}
\label{time_stretching_method}

Table \ref{table_brightness_novae} lists the 
distance modulus in the $V$ band, $(m-M)_V$, used in this work.
We derived the distance $d$ from these $(m-M)_V$-values and
checked them with the available distance estimates in the main text.
For V1500 Cyg, CP Lac, and CP Pup, we have the reliable trigonometric
distances $d= 1567^{+270}_{-192}$, $1128^{+32}_{-32}$, and
$757.3^{+8.5}_{-10.1}$ pc, respectively,
from the Gaia DR3 parallax results \citep{bai21rf}.
Our derived distances of these three novae are consistent with
the Gaia geometric distances, as listed in Table \ref{table_brightness_novae}.
It should be noted that the Gaia parallax gives a reasonable/correct
distance if the parallax error is relatively small, say $<$30\%
\citep[e.g.,][]{schaefer22}.
However, if the parallax is negative or with larger error,
its distance depends largely on the prior information of stars (Galactic
star distribution) in the Bayesian statistics calculation \citep{bai21rf}.
With keeping this caution in mind, for the other four novae V1674 Her,
V838 Her, V597 Pup, and V5589 Sgr, our derived distances are broadly
consistent with the Gaia geometric distances.
We found no Gaia source corresponding to the rest V5583 Sgr.

If there are no reliable Gaia DR3 parallax being available
(as in V1674 Her, V838 Her, V597 Pup, V5583 Sgr, and V5589 Sgr),
we are also able to calculate its distance modulus in the $V$ band
by the time-stretching method \citep{hac10k, hac20skhs}.
This is a powerful way to obtain $(m-M)_V$ toward a nova,
and has ever been applied to a number of novae
\citep{hac10k, hac14k, hac15k, hac16k, hac18kb, hac25kv392per, hac20skhs,
hac24km, hac25kw, kat25hsa}.
We use the results of the time-stretching method as a 
confirmation to the direct fit determinations of each $(m-M)_V$.
It should be noted that, in the present paper, we do not directly use
the result of time-stretching method.

Nova light curves often show a common property; if two nova light curves
are plotted in the logarithmic time and shift in the vertical and horizontal
directions, the major parts of free-free emission light curves are
overlapped independently of the WD mass, chemical composition, and speed
class of a nova \citep{hac06kb, hac20skhs}.
Using this remarkable property, we can determine the distance modulus
in the $V$ band, $(m-M)_V$, to a nova.

Here, we describe the $V$ light curves of the target nova as
$(m[t])_{V,\rm target}$ and the template nova $(m[t])_{V,\rm template}$.
When we adopt an appropriate time-stretching parameter $f_{\rm s}$,
these two nova $V$ light curves overlap each other.
We shift the template nova light curve in the horizontal direction
by a factor of $f_{\rm s}$ in the logarithmic scale
($t \rightarrow t\times f_{\rm s}$),
and move vertically down by $\Delta V$. This vertical shift
can be written as
\begin{equation}
(m[t])_{V,\rm target} = \left((m[t \times f_{\rm s}])_V
+ \Delta V\right)_{\rm template}.
\label{overlap_brigheness}
\end{equation}
As the two nova light curves overlap,
their distance moduli in the $V$ band satisfy
\begin{equation}
(m-M)_{V,\rm target} 
= ( (m-M)_V + \Delta V )_{\rm template} - 2.5 \log f_{\rm s}.
\label{distance_modulus_formula}
\end{equation}
Here, $m_V$ and $M_V$ are the apparent and absolute $V$ magnitudes,
and $(m-M)_{V, \rm target}$ and $(m-M)_{V, \rm template}$ are
the distance moduli in the $V$ band
to the target and template novae \citep{hac20skhs}, respectively.

\begin{figure*}
\gridline{\fig{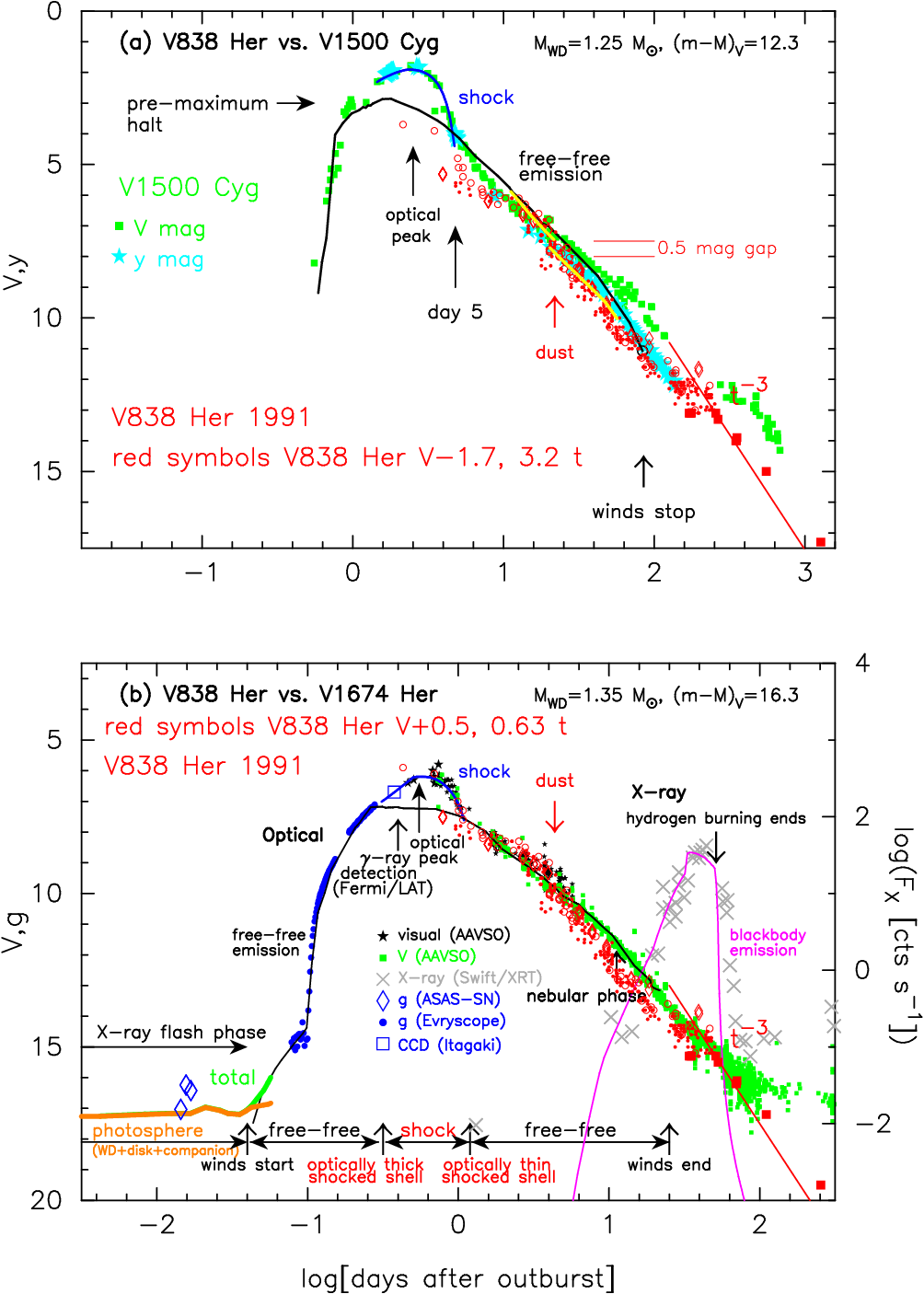}{0.35\textwidth}{}
          \fig{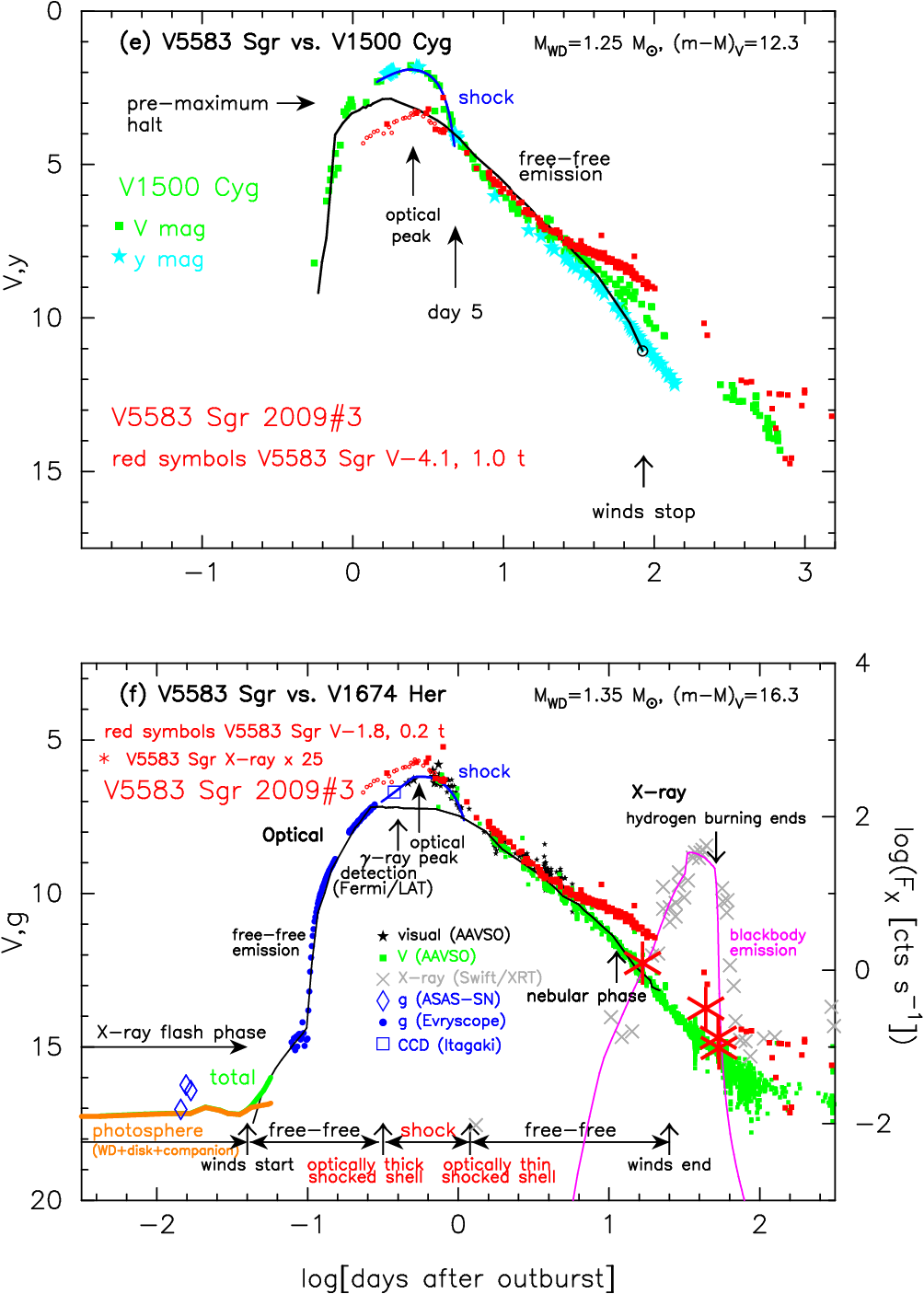}{0.35\textwidth}{}
          }
\gridline{
          \fig{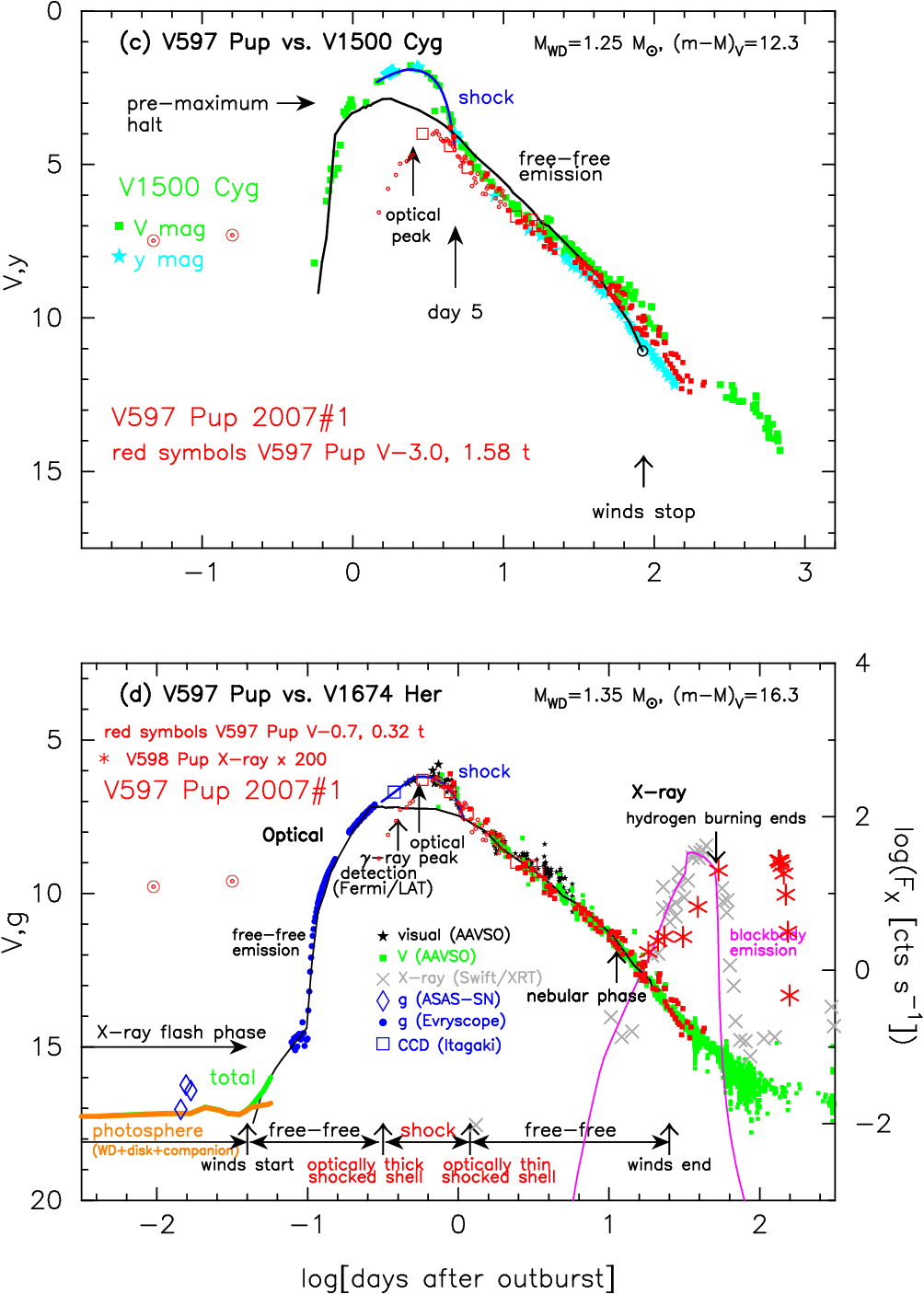}{0.35\textwidth}{}
          \fig{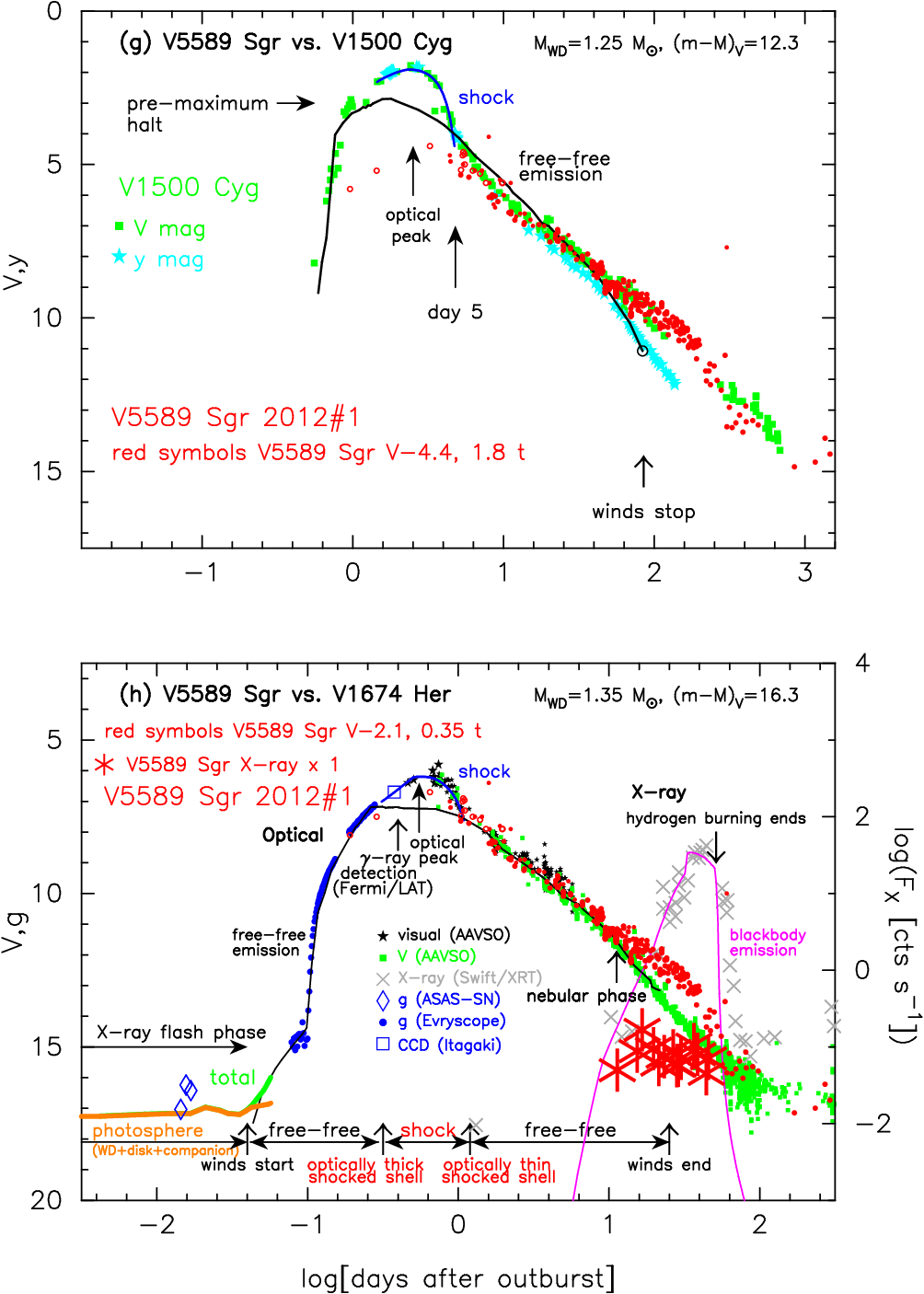}{0.35\textwidth}{}
          }
\caption{
(a) The $V$/visual light curve of V838 Her (red symbols) against
a logarithmic time, which is compared with the V1500 Cyg $V$/$y$ light
curve.  (b) V838 Her is compared with V1674 Her.
(c)(d) Same as those of (a)(b), but for V597 Pup.
(e)(f) Same as those of (a)(b), but for V5583 Sgr.
(g)(h) Same as those of (a)(b), but for V5589 Sgr.
The sources of the optical and X-ray data are shown in the main text.
\label{v838_her_v597pup_v5583sgr_v5589sgr}}
\end{figure*}

\subsection{V1674 Her 2021}
\label{distance_v1674_her}

This remarkable similarity is demonstrated in
Figure \ref{v1500cyg_1674her_model_v_observation_logscale}, which
compares the $V$ light curve of V1674 Her with V1500 Cyg.
These two novae overlap if we squeeze
the timescale of V1500 Cyg in 
Figure \ref{v1500cyg_1674her_model_v_observation_logscale}(a)
by 5 times and shift down the $V$ magnitude
by 2.3 mag as labeled ``V1500 Cyg V+2.3, 0.2 t'' in  
Figure \ref{v1500cyg_1674her_model_v_observation_logscale}(b).

It should be noted that we try to overlap the post-maximum phase 
as long/much as possible. In this phase the observed light curves 
follow free-free emission model light curves.  On the other hand,
the light curves around the optical peak are not overlapped. 
This is because the time-stretching method can
be applied only to free-free emission light curves 
\citep[see ][for the physical explanation]{hac20skhs}, 
not to the part of blackbody luminosity.

Thus, we match the epoch of rapid decay from the peak of V1674 Her
on day 1.0 with the similar rapid decay epoch of V1500 Cyg on day 5
to overlap the two $V$ light curves. 

In Figure \ref{v1500cyg_1674her_model_v_observation_logscale}(b),
we regard V1674 Her as the target and V1500 Cyg as the template
in Equation (\ref{overlap_brigheness}).
As V1500 Cyg evolves 5 times slower, we adopt $f_{\rm s}= 0.2$
and $\Delta V= +2.3\pm 0.2$ and have the relation of
\begin{equation}
(m-M)_{V, \rm V1674~Her}
= (m - M + \Delta V)_{V, \rm V1500~Cyg} - 2.5 \log 0.2
= 12.3 + 2.3\pm 0.2 + 1.75 = 16.35\pm 0.2,
\label{distance_modulus_v1674_her_lv_vul_v}
\end{equation}
where we adopt $(m-M)_{V, \rm V1500~Cyg}=12.3$ from the direct fit of
our model $V$ light curve with the observation (Table
\ref{table_brightness_novae}).
This $(m-M)_{V, \rm V1500~Cyg}=12.3$ is also calculated from
Equation (\ref{distance_modulus_extinction}) with the extinction
of $E(B-V)=0.43$ \citep{hac26kv1674her3} and
the Gaia DR3 distance \citep{bai21rf}.
This result of $(m-M)_{V, \rm V1674~Her}=16.35\pm 0.2$
is the same as that of \citet{kat25hsa}, obtained
with the same time-stretching method
but against the other three template novae,
LV Vul, V339 Del, and KT Eri.
Thus, we have the same $(m-M)_{V, \rm V1674~Her}=16.3$ as that of our 
direct fit in Section \ref{full_v1674_her_2021}, and then we obtain
$M_{V,\rm max}= -10.2$ from $m_{V,\rm max}= 6.1$ of CBET No.4977.

\subsection{V838 Her 1991}
\label{distance_v838_her}


We plot the $V$ and visual light curve of V838 Her outburst
in Figure \ref{v838_her_v597pup_v5583sgr_v5589sgr}(a) and (b),
which is the same as those in Figure 8 of
\citet{kat09v838her}.  Here, we adopt the outburst day 
of $t_{\rm OB}=$JD 2448340.0 as day zero.
In this plot, we regard V1500 Cyg as the target and V838 Her
as the template in Equation (\ref{overlap_brigheness}).
Soon after the optical maximum, V838 Her entered a very rapid
decline phase, followed by a smooth decline.
This rapid decline on day 1.6 corresponds to the rapid decline
of the optically thick shocked shell brightness (blue line) of V1500 Cyg
as shown in Figure \ref{v838_her_v597pup_v5583sgr_v5589sgr}(a),
the epoch of which is denoted by the black arrow labeled ``day 5'' for 
V1500 Cyg.
Note that the black arrows
and their descriptions indicate V1500 Cyg while the red ones are for V838 Her.
We overlap the $V$ and visual light curves of V838 Her to the $V$/$y$
light curves of V1500 Cyg
by adjusting the phases/epochs of rapid decline (day 5 in V1500 Cyg
and day 1.6 in V838 Her).
%
As V1500 Cyg evolves 3.2 times slower, we adopt $f_{\rm s}= 3.2$
($\log f_{\rm s} = 0.50$)
and $\Delta V= -1.7\pm0.2$ and have the relation of
\begin{equation}
(m-M)_{V, \rm V1500~Cyg} 
= (m - M)_{V, \rm V838~Her} -1.7\pm0.2 - 2.5 \log 3.2.
\label{distance_modulus_v838_her_v1500_cyg_v}
\end{equation}
Substituting $(m-M)_{V, \rm V1500~Cyg}=12.3$ into Equation
(\ref{distance_modulus_v838_her_v1500_cyg_v}),
we obtain $(m-M)_{V, \rm V838~Her}=15.25\pm 0.2$.

In Figure \ref{v838_her_v597pup_v5583sgr_v5589sgr}(b),
we regard V1674 Her as the target and V838 Her as the template
in Equation (\ref{overlap_brigheness}).  Here, we also adjust the
phases/epochs of rapid decline (day 1 in V1674 Her and day 1.6 in V838 Her).
As V1674 Her evolves 0.63 times slower (or 1.6 times faster),
we adopt $f_{\rm s}= 0.63$ ($\log f_{\rm s} = -0.20$)
and $\Delta V= +0.5\pm0.2$ and have the relation of
\begin{equation}
(m-M)_{V, \rm V1674~Her}
= (m - M)_{V, \rm V838~Her} +0.5\pm0.2 - 2.5 \log 0.63.
\label{distance_modulus_v838_her_v1674_her_v}
\end{equation}
Substituting $(m-M)_{V, \rm V1674~Her}=16.3$ into Equation
(\ref{distance_modulus_v838_her_v1674_her_v}),
we obtain $(m-M)_{V, \rm V838~Her}=15.3\pm 0.2$.
Thus, we have the same $(m-M)_{V, \rm V838~Her}=15.3$ as that of our
direct fit in Section \ref{full_v838_her_1991}, and then obtain
$M_{V,\rm max}= -10.0$ from $m_{V,\rm max}= 5.3$ \citep[Table 6 of][]{ozd18}.

\subsection{V597 Pup 2007\#1}
\label{distance_v597_pup}


We plot the $V$, visual, and SMEI light curves of V597 Pup in Figure
\ref{v838_her_v597pup_v5583sgr_v5589sgr}(c)and (d)
together with the light curves of (c) V1500 Cyg and (d) V1674 Her.
The data of V597 Pup are
denoted by red symbols.
The data of V597 Pup are the same as Figure 20 of \citet{hac10k}, which
are taken from SMARTS \citep{wal12bt},
AAVSO, VSOLJ, and IAUC No.8895 and No.8902.
We add the SMEI magnitudes \citep[small open red circles, ][]{hou16db}.
Excluding the first two SMEI data points (small open red circles
encircled by a larger red circle), we adopt the outburst day to be
$t_{\rm OB}=$ JD 2454416.9 $=$ UT 2007 November 13.4.

In Figure \ref{v838_her_v597pup_v5583sgr_v5589sgr}(c),
as V1500 Cyg evolves 1.58 times slower than V597 Pup,
we adopt $f_{\rm s}= 1.58$ ($\log f_{\rm s}= 0.2$) 
and $\Delta V= -3.0\pm0.2$ and have the relation of
\begin{equation}
(m-M)_{V, \rm V1500~Cyg}
= (m - M)_{V, \rm V597~Pup} -3.0\pm0.2 - 2.5 \log 1.58.
\label{distance_modulus_v597_pup_v1500_cyg_v}
\end{equation}
Substituting $(m-M)_{V, \rm V1500~Cyg}=12.3$ into Equation
(\ref{distance_modulus_v597_pup_v1500_cyg_v}),
we obtain $(m-M)_{V, \rm V597~Pup}=15.8\pm 0.2$.

In Figure \ref{v838_her_v597pup_v5583sgr_v5589sgr}(d),
we regard V1674 Her as the target and V597 Pup as the template
in Equation (\ref{overlap_brigheness}).
Here, we adjust the rise epoch of the supersoft X-ray flux
(red asterisks) of V597 Pup with that of V1674 Her (gray crosses).
As V1674 Her evolves 0.32 times slower (or 3.2 times faster),
we adopt $f_{\rm s}= 0.32$ ($\log f_{\rm s}= -0.5$)
and $\Delta V= -0.7\pm0.2$ and have the relation of
\begin{equation}
(m-M)_{V, \rm V1674~Her}
= (m - M)_{V, \rm V597~Pup} -0.7\pm0.2 - 2.5 \log 0.32.
\label{distance_modulus_v597_pup_v1674_her_v}
\end{equation}
Substituting $(m-M)_{V, \rm V1674~Her}=16.3$ into Equation
(\ref{distance_modulus_v597_pup_v1674_her_v}),
we obtain $(m-M)_{V, \rm V597~Pup}=15.75\pm 0.2$.
Thus, we have the same $(m-M)_{V, \rm V597~Pup}=15.8$ as that of our
direct fit in Section \ref{full_v597_pup_2007}, and then calculate
$M_{V,\rm max}= -9.0$ from $m_{V,\rm max}= 6.8$ in the VSOLJ data.


\subsection{V5583 Sgr 2009\#3}
\label{distance_v5583_sgr}


We plot the $V$, visual, and SMEI light curves of V5583 Sgr 
in Figure \ref{v838_her_v597pup_v5583sgr_v5589sgr}(e) and (f),
together with (e) V1500 Cyg and (f) V1674 Her.
The data of V5583 Sgr are the same as those in Figure 67 of \citet{hac19kb},
which are taken from SMARTS \citep{wal12bt}, AAVSO, VSOLJ, and IAUC No.9061.
We add the SMEI magnitudes \citep[small open red circles, ][]{hou16db},
but shift them down by 0.4 mag to match with the $V$ brightness.
Here, we assume the outburst day of V5583 Sgr to be 
$t_{\rm OB}=(t=0=)$JD 2455048.0 (UT 2009 August 4.5).  
We adopt \citet{hou16db}'s SMEI data from the first data on MJD 55048.67
upto the data on MJD 55051 (JD 2455051.5),
because the SMEI light curve seems to level off for a while
and is inconsistent with the other 
photometry such as the Solar TErrestrial RElations Observatory (STEREO)
Heliospheric Imager \citep{holds14rb}.

In Figure \ref{v838_her_v597pup_v5583sgr_v5589sgr}(e),
we regard V1500 Cyg as the target and V5583 Sgr as the template
in Equation (\ref{overlap_brigheness}).  Because the SMEI band width
is wider than that of the $V$ band and the SMEI magnitudes
are systematically 0.4 mag brighter than the other $V$ data,
we shift down the SMEI data by 0.4 mag and overlap them with the
other $V$ brightnesses as much as possible.
As V1500 Cyg evolves as fast as V5583 Sgr,
we adopt $f_{\rm s}= 1.0$ ($\log f_{\rm s}= 0.0$)
and $\Delta V= -4.1\pm0.2$ and have the relation of
\begin{equation}
(m-M)_{V, \rm V1500~Cyg}
= (m - M)_{V, \rm V5583~Sgr} -4.1\pm0.2 - 2.5 \log 1.0.
\label{distance_modulus_v5583_sgr_v1500_cyg_v}
\end{equation}
Substituting $(m-M)_{V, \rm V1500~Cyg}=12.3$ into Equation
(\ref{distance_modulus_v5583_sgr_v1500_cyg_v}),
we obtain $(m-M)_{V, \rm V5583~Sgr}=16.4\pm 0.2$.

In Figure \ref{v838_her_v597pup_v5583sgr_v5589sgr}(f),
we regard V1674 Her as the target and V5583 Sgr as the template
in Equation (\ref{overlap_brigheness}).
It should be noted that we overlap the X-ray light curve of V5583 Sgr
(large red asterisks) with that of V1674 Her (gray crosses and solid magenta
line) as much as possible.
As V1674 Her evolves 0.2 times slower (or 5.0 times faster) than V5583 Sgr,
we adopt $f_{\rm s}= 0.2$ ($\log f_{\rm s}= -0.7$)
and $\Delta V= -1.2\pm0.2$ and have the relation of
\begin{equation}
(m-M)_{V, \rm V1674~Her}
= (m - M)_{V, \rm V5583~Sgr} -1.8\pm0.2 - 2.5 \log 0.2.
\label{distance_modulus_v5583_sgr_v1674_her_v}
\end{equation}
Substituting $(m-M)_{V, \rm V1674~Her}=16.3$ into Equation
(\ref{distance_modulus_v5583_sgr_v1674_her_v}),
we obtain $(m-M)_{V, \rm V5583~Sgr}=16.35\pm 0.2$.
Thus, we have the same value of $(m-M)_{V, \rm V5583~Sgr}=16.4$ as
that of our direct fit in Section \ref{full_v5583_sgr_2009_no3}, both from 
Figures \ref{v838_her_v597pup_v5583sgr_v5589sgr}(e) and (f), and then
obtain $M_{V,\rm max}= -9.0$ from $m_{V,\rm max}= 7.4$ \citep{maehara09}.

\subsection{V5589 Sgr 2012\#1}
\label{distance_v5589_sgr}


Figure \ref{v838_her_v597pup_v5583sgr_v5589sgr}(g) shows
the $V$ and visual light curves of V5589 Sgr together with the $V$/$y$
light curves of V1500 Cyg.  
The data of V5589 Sgr are the same as those in Figure 91 of \citet{hac19kb},
which are taken from SMARTS \citep{wal12bt},
AAVSO, VSOLJ, and CBET No.3089.
Here, we assume the outburst day of V5589 Sgr
to be $t_{\rm OB}=$ JD 2456038.0 (UT 2012 April 20.5).   
Here, we regard V1500 Cyg as the target and V5589 Sgr as the template
in Equation (\ref{overlap_brigheness}).
As V1500 Cyg evolves 1.8 times slower (0.56 times faster) than V5589 Sgr,
we adopt $f_{\rm s}= 1.8$ ($\log f_{\rm s}= 0.25$)
and $\Delta V= -4.4\pm 0.2$ and have the relation of
\begin{equation}
(m-M)_{V, \rm V1500~Cyg}
= (m - M)_{V, \rm V5589~Sgr} -4.4\pm0.2 - 2.5 \log 1.8.
\label{distance_modulus_v5589_sgr_v1500_cyg_v}
\end{equation}
Substituting $(m-M)_{V, \rm V1500~Cyg}=12.3$ into Equation
(\ref{distance_modulus_v5589_sgr_v1500_cyg_v}),
we obtain $(m-M)_{V, \rm V5589~Sgr}=17.3\pm 0.2$.

In Figure \ref{v838_her_v597pup_v5583sgr_v5589sgr}(h),
we regard V1674 Her as the target and V5589 Sgr as the template
in Equation (\ref{overlap_brigheness}).
Here we try to fit the supersoft X-ray phase of two objects
as much as possible.
We suppose that the soft X-rays of V5589 Sgr are heavily absorbed
because of a large extinction of $E(B-V)=0.865$.
As V1674 Her evolves 0.35 times slower (2.8 times faster) than V5589 Sgr,
we adopt $f_{\rm s}= 0.35$ ($\log 0.35= -0.45$)
and $\Delta V= -2.1\pm0.2$ and have the relation of
\begin{equation}
(m-M)_{V, \rm V1674~Her}
= (m - M)_{V, \rm V5589~Sgr} -2.1\pm0.2 - 2.5 \log 0.35.
\label{distance_modulus_v5589_sgr_v1674_her_v}
\end{equation}
Substituting $(m-M)_{V, \rm V1674~Her}=16.3$ into Equation
(\ref{distance_modulus_v5589_sgr_v1674_her_v}),
we obtain $(m-M)_{V, \rm V5589~Sgr}=17.3\pm 0.2$.
Thus, we have the same $(m-M)_{V, \rm V5589~Sgr}=17.3$ as that of
our direct fit in Section \ref{full_v5589_sgr_2012_no1}, and then obtain 
$M_{V,\rm max}= -8.5$ from $m_{V,\rm max}= 8.8$ of CBET No.3089.
These results are summarized in Table \ref{table_brightness_novae}.

\end{document}